%% file: main.tex
\documentclass[11pt]{article}
\usepackage[utf8]{inputenc}

\input{header.tex}
\input{macro.tex}

\title{Dynamic $(1+\epsilon)$-Approximate Matching Size\\in Truly Sublinear Update Time}

\author{Sayan Bhattacharya\thanks{University of Warwick \texttt{s.bhattacharya@warwick.ac.uk}} \thanks{Supported by Engineering and Physical Sciences Research Council, UK (EPSRC) Grant EP/S03353X/1.} \and Peter Kiss\thanks{University of Warwick, Max-Planck-Institut für Informatik \texttt{peter.kiss@warwick.ac.uk}} \and Thatchaphol Saranurak\thanks{University of Michigan \texttt{thsa@umich.edu}} \thanks{Supported by NSF CAREER grant 2238138.}} 

\date{}

\begin{document}

\maketitle

\pagenumbering{gobble}
\input{0-abstract.tex}
\pagebreak



\pagebreak

\tableofcontents

\pagebreak

\pagenumbering{arabic}
\input{1-introduction.tex}
\input{2-overview.tex}
\input{3-prelim.tex}
\input{4-matching-oracle.tex}

\input{5-boosting.tex}

\input{6-sublinear.tex}
\input{7-dynamic.tex}

\bibliographystyle{alpha}
\bibliography{ref-abb,ref}

\appendix
\input{A-simple-oracle.tex}

\input{B-vertex-sparsification.tex}
\input{C-trivial-algo.tex}
\input{table.tex}

\end{document}

%% file: header.tex
\usepackage[utf8]{inputenc}
\usepackage[T1]{fontenc}

\usepackage{geometry}
\usepackage[bottom]{footmisc}
\usepackage{todonotes}
\usepackage{xspace}
\usepackage{array}
\usepackage{verbatim}
\usepackage{float}
\usepackage{multirow}
\usepackage{comment}

\usepackage{nicefrac}
\usepackage{algorithm}
\usepackage{algorithmicx}
\usepackage[noend]{algpseudocode}

\usepackage{amssymb}
\usepackage{amsmath}
\usepackage{amsthm}
\usepackage{dsfont}
\usepackage{footnote}

\usepackage{xcolor}
\usepackage{nameref}
\definecolor{ForestGreen}{rgb}{0.1333,0.5451,0.1333}
\definecolor{DarkRed}{rgb}{0.65,0,0}
\definecolor{Red}{rgb}{1,0,0}
\usepackage[linktocpage=true,
pagebackref=true,colorlinks,
linkcolor=DarkRed,citecolor=ForestGreen,
bookmarks,bookmarksopen,bookmarksnumbered]
{hyperref}

\usepackage{cleveref}

\usepackage{thm-restate}
\usepackage[tableposition=bottom]{caption}
\usepackage[all]{hypcap}
\usepackage{subcaption}
\usepackage{framed}
\usepackage[framemethod=tikz]{mdframed}
\usepackage[skins]{tcolorbox}
\usepackage{enumerate}
\makeatletter
\g@addto@macro{\maketitle}{\@thanks}
\makeatother

\newtheorem{thm}{Theorem}[section]
\newtheorem{cor}[thm]{Corollary}
\newtheorem{prop}[thm]{Proposition}

\newtheorem{invariant}[thm]{Invariant}
\newtheorem{lem}[thm]{Lemma}
\newtheorem{definition}[thm]{Definition}
\newtheorem{obs}[thm]{Observation}
\newtheorem{claim}[thm]{Claim}
\newtheorem{question}[thm]{Question}

\renewcommand{\algorithmiccomment}[1]{\bgroup\hfill$\rhd$~#1\egroup}

\newcounter{note}[section]

\newenvironment{wrapper}[1]
{
	\begin{center}
		\begin{minipage}{\linewidth}
			\begin{mdframed}[hidealllines=true, backgroundcolor=gray!20, leftmargin=0cm,innerleftmargin=0.4cm,innerrightmargin=0.4cm,innertopmargin=0.4cm,innerbottommargin=0.4cm,roundcorner=10pt]
				#1}
			{\end{mdframed}
		\end{minipage}
	\end{center}
} 

\renewcommand{\paragraph}[1]{\medskip\noindent\textbf{#1}}

\ifdefined\ShowComment
    \def\thatchaphol#1{\marginpar{$\leftarrow$\fbox{T}}\footnote{$\Rightarrow$~{\sf\textcolor{ForestGreen}{#1 --Thatchaphol}}}}   \def\sayan#1{\marginpar{$\leftarrow$\fbox{S}}\footnote{$\Rightarrow$~{\sf\textcolor{purple}{#1 --Sayan}}}}  \def\peter#1{\marginpar{$\leftarrow$\fbox{P}}\footnote{$\Rightarrow$~{\sf\textcolor{green}{#1 --Peter}}}}
\else
    \def\thatchaphol#1{}    
    \def\sayan#1{}
    \def\peter#1{}
\fi 

\usepackage{etoolbox}
\usepackage{tikz}
\usetikzlibrary{tikzmark}
\usetikzlibrary{calc}

\errorcontextlines\maxdimen

\newcommand{\ALGtikzmarkcolor}{black}
\newcommand{\ALGtikzmarkextraindent}{4pt}
\newcommand{\ALGtikzmarkverticaloffsetstart}{-.5ex}
\newcommand{\ALGtikzmarkverticaloffsetend}{-.5ex}
\makeatletter
\newcounter{ALG@tikzmark@tempcnta}

\newcommand\ALG@tikzmark@start{%
	\global\let\ALG@tikzmark@last\ALG@tikzmark@starttext%
	\expandafter\edef\csname ALG@tikzmark@\theALG@nested\endcsname{\theALG@tikzmark@tempcnta}%
	\tikzmark{ALG@tikzmark@start@\csname ALG@tikzmark@\theALG@nested\endcsname}%
	\addtocounter{ALG@tikzmark@tempcnta}{1}%
}

\def\ALG@tikzmark@starttext{start}
\newcommand\ALG@tikzmark@end{%
	\ifx\ALG@tikzmark@last\ALG@tikzmark@starttext
	\else
	\tikzmark{ALG@tikzmark@end@\csname ALG@tikzmark@\theALG@nested\endcsname}%
	\tikz[overlay,remember picture] \draw[\ALGtikzmarkcolor] let \p{S}=($(pic cs:ALG@tikzmark@start@\csname ALG@tikzmark@\theALG@nested\endcsname)+(\ALGtikzmarkextraindent,\ALGtikzmarkverticaloffsetstart)$), \p{E}=($(pic cs:ALG@tikzmark@end@\csname ALG@tikzmark@\theALG@nested\endcsname)+(\ALGtikzmarkextraindent,\ALGtikzmarkverticaloffsetend)$) in (\x{S},\y{S})--(\x{S},\y{E});%
	\fi
	\gdef\ALG@tikzmark@last{end}%
}

\apptocmd{\ALG@beginblock}{\ALG@tikzmark@start}{}{\errmessage{failed to patch}}
\pretocmd{\ALG@endblock}{\ALG@tikzmark@end}{}{\errmessage{failed to patch}}
\makeatother

%% file: macro.tex
\global\long\def\eps{\epsilon}
\global\long\def\Otil{\tilde{O}}
\global\long\def\mem{\mathtt{mem}}
\global\long\def\match{\mathtt{match}}
\global\long\def\poly{\mathrm{poly}}
\global\long\def\polylog{\mathrm{polylog}}
\global\long\def\lowdeg{\mathrm{low}}
\global\long\def\sizeIn{\delta_{\mathrm{in}}}
\global\long\def\sizeOut{\delta_{\mathrm{out}}}
\global\long\def\lca{\textsc{lca}}
\global\long\def\E{\mathbb{E}}
\global\long\def\Mhat{\widehat{M}}
\global\long\def\Gbar{\overline{G}}
\global\long\def\vol{\mathrm{vol}}
\global\long\def\sp{\mathrm{sp}}

\global\long\def\testperm{\textsc{TestPerm}}
\global\long\def\VO{\mathsf{VO}}
\global\long\def\EO{\mathsf{EO}}
\global\long\def\TRUE{\mathsf{TRUE}}
\global\long\def\FALSE{\mathsf{FALSE}}
\global\long\def\dbar{\overline{d}}

\newcommand{\calA}{\mathcal{A}}

\renewcommand{\Otil}{\tilde{O}}

\renewcommand{\E}{\mathbb{E}}%

\renewcommand{\eps}{\epsilon}%
\newcommand{\p}{\textsc{P}}%
\renewcommand{\poly}{\mathrm{poly}}
\renewcommand{\polylog}{\mathrm{polylog}}%

 \newcommand{\algo}{\texttt{Augment}}
 \newcommand{\largematch}{\texttt{LargeMatching}} 
 \newcommand{\stack}{\texttt{stack}}
 \newcommand{\mate}{\texttt{mate}} 
\newcommand{\inp}{\texttt{in}}
\newcommand{\out}{\texttt{out}}

\global\long\def\alg{\texttt{Alg}}
\global\long\def\largematchoracle{\texttt{LargeMatchingOracle}}

%% file: 0-abstract.tex
\begin{abstract}
We show a fully dynamic algorithm for maintaining $(1+\epsilon)$-approximate \emph{size} of maximum matching of the graph with $n$ vertices and $m$ edges using $m^{0.5-\Omega_{\epsilon}(1)}$ update time. This is the first polynomial improvement over the long-standing $O(n)$ update time, which can be trivially obtained by periodic recomputation. Thus, we resolve the value version of a major open question of the dynamic graph algorithms literature (see, e.g., ~{[}Gupta and Peng FOCS'13{]}, {[}Bernstein and Stein SODA'16{]}, {[}Behnezhad and Khanna SODA'22{]}).


Our key technical component is the first sublinear algorithm for $(1,\epsilon n)$-approximate maximum matching with sublinear running time on \emph{dense} graphs. 
All previous algorithms suffered a multiplicative approximation factor of at least $1.499$ or assumed that the graph has a very small maximum degree.

\end{abstract}

%% file: 1-introduction.tex
\section{Introduction}
\label{sec:intro}

We study the dynamic version of the maximum matching problem, a cornerstone of combinatorial optimization \cite{kuhn1955hungarian,edmonds1965paths,edmonds1965maximum}. In the \emph{dynamic matching} problem, the task is to build a data structure that, given a graph $G$ with $n$ vertices and $m$ edges undergoing both edge insertions and deletions, maintains an (approximate) maximum matching of $G$ or, in the value version, just the size of the maximum matching, denoted by $\mu(G)$. The goal is to minimize the \emph{update time} required to update the solution after each edge change. 

The first non-trivial algorithm for this problem was by Sankowski \cite{sankowski2007faster} 15 years ago, which exactly maintains the maximum matching \emph{size }using $O(n^{1.495})$ update time, which is recently improved to $O(n^{1.407})$ \cite{van2019dynamic}. Unfortunately, this latter bound is tight under the \emph{hinted OMv} conjecture \cite{van2019dynamic}. Furthermore, in sparse graphs, even $m^{1-o(1)}$ update time is required assuming the $k$-cycle conjecture \cite{probst2020new}. These strong conditional lower bounds have shifted the attention of researchers to \emph{approximate} matching. An $\alpha$-approximate matching is a matching of size at least $\mu(G)/\alpha$. The following has become one of the holy-grail questions in the dynamic graph algorithms and fine-grained complexity communities \cite{abboud2017distributed}:
\begin{question}
\label{que:holy}Is there a dynamic $(1+\eps)$-approximate matching algorithm with polylogarithmic update time for an arbitrarily small constant $\eps$? 
\end{question}

The current state of the art is, however, still very far from this goal. A straightforward algorithm with $O(n)$ amortized update time is to simply recompute a $(1+\eps)$-approximate matching from scratch in $O(m)$ time \cite{duan2014linear} every after $2\eps m/n$ edge updates.\footnote{See \Cref{sec:trivial} for the proof of this simple algorithm.} Surprisingly, this easy $O(n)$ bound already captures the limitation of all known techniques! An improved algorithm with $O(\sqrt{m})$ update time was given ten years ago by Gupta and Peng \cite{gupta2013fully}, but it still takes $O(n)$ time in dense graphs. Very recently, Assadi et al.~\cite{assadi2022regularity} showed how to obtain $O(n/(\log^{*}n)^{\Omega(1)})$ update time, but their regularity-lemma-based approach inherently cannot give an improvement larger than a $2^{\Theta(\sqrt{\log n})}=n^{o(1)}$ factor. Until now, no dynamic $(1+\eps)$-approximate algorithms can break through the naive $O(n)$ barrier by a polynomial factor.\footnote{In contrast, in the \emph{partially} dynamic setting where graphs undergo edge insertions only or edge deletions only, there are many algorithms with polylogarithmic amortized update time \cite{gupta2014maintaining,grandoni20191+,bernstein2020deterministic,jambulapati2022regularized,bhattacharya2023dynamic}.} 

Intensive research during the last decade instead showed how to speed up update time by relaxing the approximation factor. The influential work by Onak and Rubinfeld \cite{onak2010maintaining} gave the first dynamic matching algorithm with polylogarithmic update time that maintains a large constant approximate maximum matching. Then, Baswana, Gupta and Sen \cite{baswana2011fully} showed a dynamic maximal matching with logarithmic update time, which gives $2$-approximation. A large body of work then refined this result in various directions, including constant update time \cite{solomon2016fully}, deamortization \cite{charikar2018fully,bernstein2019deamortization,kiss2022improving}, and derandomization \cite{bhattacharya2016new,arar2018dynamic,bhattacharya2019deterministically,wajc2020rounding,bhattacharya2021deterministic}. In 2015, Bernstein and Stein \cite{bernstein2015fully,bernstein2016faster} showed a novel approach for maintaining a $(3/2+\eps)$-approximate matching using $\Otil(m^{1/4})=\Otil(\sqrt{n})$ update time.\footnote{We use $\Otil(\cdot)$ to hide $\polylog(n)$ factor throughout the paper.} Refinement of this approach and new trade-off results with approximation in the range $(3/2,2)$ were also intensively studied \cite{behnezhad2020fully,grandoni2022maintaining,kiss2022improving,behnezhad2022new,roghani2022beating}. All these techniques, however, seem to get stuck at $(3/2)$-approximation. 

Very recently, the above long-standing trade-off was improved by Behnezhad \cite{behnezhad2023dynamic} and, independently, by Bhattacharya et al.~\cite{bhattacharya2023dynamic} via a new connection to sublinear and streaming algorithms. To maintain maximum matching \emph{size}, they gave $1.973$-approximation algorithms with polylogarithmic update time, and, on bipartite graphs, Behnezhad \cite{behnezhad2023dynamic} pushed it further to $(3/2-\Omega(1))$-approximation in $\Otil(\sqrt{n})$ update time. While this new connection is very inspiring, it has been a key open problem \cite{behnezhad2023beating} whether non-trivial $(1+\eps)$-approximate matching algorithms in dense graphs exist in the sublinear model. Hence, it remains unclear whether an improved dynamic $(1+\eps)$-approximation algorithm is possible via this new connection or even possible at all.

Indeed, in this paper, we give the first dynamic $(1+\eps)$-approximate matching \emph{size} algorithm that finally improves the $O(n)$ bound by a polynomial factor, formally stated below.
\begin{restatable}{thm}{maindynamic}\label{thm:main dynamic}
There is a dynamic $(1+\eps)$-approximate matching \emph{size} algorithm with $m^{0.5-\Omega_{\epsilon}(1)}$ worst-case update time. 

The algorithm is randomized and works against an adaptive adversary with high probability. Moreover, the algorithm maintains $(1+\epsilon)$-approximate matching $M$ of $G$ in the sense that, given a vertex $v$, it can return a matched edge $(v,v')\in M$ or $\bot$ if $v\notin V(M)$ in $m^{0.5+f(\epsilon)}$ time, where $f$ is an increasing function such that $f(\epsilon) \rightarrow 0$ when $\epsilon\rightarrow 0$.
\end{restatable}

It has been asked repeatedly \cite{gupta2013fully,bernstein2015fully,bernstein2016faster} whether there exists a dynamic $(1+\eps)$-approximate matching algorithm with $m^{0.5-\Omega_{\eps}(1)}$ update time. \Cref{thm:main dynamic} thus gives an affirmative answer to the value version of this open question. Although the matching is not explicitly maintained in \Cref{thm:main dynamic}, it still supports queries whether a vertex is matched or not. The recent algorithms that only maintain the estimate of $\mu(G)$ by \cite{behnezhad2023dynamic,bhattacharya2023dynamic} inherently cannot support this query.

We obtain \Cref{thm:main dynamic} by making progress in sublinear algorithms: we show the first sublinear $(1,\eps n)$-approximate matching algorithm with truly sublinear time even in dense graphs. Here, an $(\alpha,\beta)$-approximate matching means a matching of size at least $\mu(G)/\alpha-\beta$. Given our new sublinear matching algorithm summarized below, \Cref{thm:main dynamic} follows using known techniques. 

\begin{restatable}{thm}{mainsublinear}\label{thm:main sublinear}
There is a randomized algorithm that, given the adjacency matrix of a graph $G$, in time $n^{2-\Omega_{\epsilon}(1)}$ computes with high probability a $(1,\eps n)$-approximation $\tilde{\mu}$ of $\mu(G)$. 

After that, given a vertex $v$, the algorithm returns in $n^{1+f(\epsilon)}$ time an edge $(v,v')\in M$ or $\bot$ if $v\notin V(M)$ where $M$ is a fixed $(1,\epsilon n)$-approximate matching, where $f$ is an increasing function such that $f(\epsilon) \rightarrow 0$ when $\epsilon\rightarrow 0$. 
\end{restatable}

We note that the additive approximation factor in \Cref{thm:main sublinear} is unavoidable for sublinear algorithms with access to only the adjacency matrix: checking whether there is zero or one edge requires $\Omega(n^2)$ adjacency matrix queries. 

Behnezhad et al.~\cite{behnezhad2023beating} posted an open question about sublinear matching algorithms as follows ``ruling out say a $1.01$-approximation in $n^{2-\Omega(1)}$ time would also be extremely interesting.''\footnote{In \cite{roghani2022beating}, they use different notation and write $0.99$-approximation instead of $1.01$.}.
Since the additive approximation factor is unavoidable for algorithms using the adjacency matrix only, the analogous question becomes whether one can rule out a $(1,n/100)$-approximation in $n^{2-\Omega(1)}$ time. 
\Cref{thm:main sublinear} answers this question negatively since we can get arbitrarily good additive approximation in $n^{2-\Omega(1)}$ time.

To put \Cref{thm:main sublinear} into the larger context of sublinear matching literature, let us discuss its history below. We  use $\Delta$ and $d$ to denote the maximum and average degree of the graph respectively. 

\paragraph{Approximating $\mu(G)$.}
One of the main goals in this area, initiated by Parnas and Ron \cite{parnas2007approximating}, is to approximate the \emph{size} of maximum matching $\mu(G)$ in sublinear time when given access to the adjacency list and matrix of an input graph. Early research on this topic focused on obtaining $O(1)$ time algorithms when $\Delta=O(1)$. However, these early work \cite{parnas2007approximating,nguyen2008constant,yoshida2012improved} may require $\Omega(n^{2})$ time on general graphs. This drawback was first addressed in \cite{kapralov2020space} and \cite{chen2020sublinear} (based on \cite{onak2012near}), both of which were then subsumed by the algorithms of Behnezhad \cite{behnezhad2022time} that compute a $(2,o(n))$-approximation in $\Otil(d+1)$ time. His algorithms are near-optimal and settle the problem in the regime of approximation ratio at least $2$. 

Subsequent work focuses on optimizing the approximation ratio within $n^{2-\Omega(1)}$ time. To compare with \Cref{thm:main sublinear}, let us discuss only  algorithms that use the adjacency matrix.
Behnezhad et al.~\cite{behnezhad2023beating} first broke the $2$-approximation barrier by computing a $(2-\Omega_{\gamma}(1),o(n))$-approximate matching in $\Otil(n^{1+\gamma})$ time. Then $(3/2,\eps n)$-approximation algorithms with $n^{2-\Theta(\eps^{2})}$ time were shown independently in \cite{bhattacharya2022sublinear,behnezhad2022sublinear}. Behnezhad et al.~\cite{behnezhad2022sublinear} improved this further to $(3/2-\Omega(1),o(n))$-approximation in $n^{2-\Omega(1)}$ time on bipartite graphs.\footnote{\cite{behnezhad2022sublinear} also announced a $\Omega(n^{1.2})$-time lower bound for $(3/2-\Omega(1),o(n))$-approximation.}
We summarize the previous work in \Cref{tab:size}.

By the first part of \Cref{thm:main sublinear}, we show that even $(1, \eps n)$-approximation is possible in $n^{2-\Omega_{\eps}(1)}$ time. 
As we mentioned, this result addresses the open question of \cite{roghani2022beating}. It remains very interesting to see an optimal approximation-time trade-off for this problem.

\paragraph{Matching Oracles.}
In the area of \emph{local computation algorithms} (LCA), initiated by Robinfeld et al.~\cite{rubinfeld2011fast,alon2012space}, we want a \emph{matching oracle }for some fixed approximate matching $M$ such that, given any vertex $v$, return $(v,v')\in M$ or $\bot$ if $v\notin V(M)$. The goal is to optimize the approximation ratio of $M$ and minimize the worst-case query time over \emph{all} vertices. Note that, given a matching oracle for an $\alpha$-approximate matching, we can compute $(\alpha,\eps n)$-approximation of $\mu(G)$ by simply querying the oracle at $O(1/\eps^{2})$ random vertices. So this is stronger than the previous goal.

The worst-case guarantee over all vertices is stronger than the expected query time for \emph{each} vertex \cite{nguyen2008constant} or for just a \emph{random} vertex \cite{yoshida2012improved,behnezhad2022time}, which is even weaker. This strong guarantee is useful for bounding the query time of \emph{adaptive queries}, which depend on answers of the previous queries, and is crucial in some applications \cite{lange2022properly}. Our approach for ``boosting'' the approximation ratio also requires adaptive queries and hence needs worst-case guarantees. 

A long line of work \cite{rubinfeld2011fast,alon2012space,reingold2016new,levi2015local,ghaffari2016improved,ghaffari2019sparsifying,ghaffari2022local} focused on building an oracle for maximal independent sets (which implies a $2$-approximate matching oracle) and culminated in an oracle by Ghaffari \cite{ghaffari2022local} that uses $\poly(\Delta\log n)$ query time with high probability. Levi et al.~\cite{levi2015local} also a showed $(1+\eps)$-approximate matching oracle with $\Delta^{O(1/\eps^{2})}\polylog(n)$ query complexity. However, all these algorithms are not sublinear in dense graphs. In this regime, the only non-trivial matching oracle was by Kapralov et al.~\cite{kapralov2020space} and has $\Otil(\Delta)$ query time, but the approximation ratio is only a large constant and is in expectation. We summarize the previous work in \Cref{tab:LCA}.

The second part of \Cref{thm:main sublinear} gives the first non-trivial matching oracle on dense graphs whose multiplicative approximation ratio is a small constant, which is $1$ in our case, but we need to pay additive approximation factor.

\paragraph{Summary. }
Our main result, \Cref{thm:main dynamic}, is the first dynamic $(1+\eps)$-approximate matching size algorithm with $m^{0.5-\Omega_{\eps}(1)}$ update time, breaking through the naive yet long-standing $O(n)$ barrier by a polynomial factor. Our key technical component, \Cref{thm:main sublinear}, makes progress in the area of sublinear-time matching algorithms on \emph{dense} graphs. Among algorithms for approximating $\mu(G)$ only, we improve the best approximation ratio from $(3/2-\Omega(1),o(n))$ by \cite{behnezhad2022sublinear} to $(1, \eps n)$. Among LCAs, it is the first one on dense graphs whose multiplicative approximation is a small constant.

\paragraph{Organization.}
First, we give an overview of our algorithms in \Cref{sec:overview}.
Then, we set up notations and give preliminaries in \Cref{sec:prelim}. 
In \Cref{sec:oracle induced}, we present a key building block which is a matching oracle for an induced graph $G[A]$ where $A$ is unknown to us. 
Using this, we show in \Cref{sec:boost} how to boost the approximation ratio of any matching oracle. By repeatedly boosting the approximation ratio, we give a $(1,\epsilon n)$-approximate matching oracle (\Cref{thm:main sublinear}) in  \Cref{sec:sublinear:main}. 
Finally, we combine this oracle with known techniques in dynamic algorithms to \Cref{thm:main dynamic} in \Cref{sec:dynamic}. 

\paragraph{Acknowledgement}

We would like to thank David Wajc for suggesting the implementation of McGregor's algorithm in the sub-linear model.

%% file: 2-overview.tex
\section{Technical Overview}
\label{sec:overview}

Our high-level approach is based on the interconnection between dynamic, sublinear, and streaming algorithms. This connection differs from the ones used in the recent results of \cite{bhattacharya2023dynamic,behnezhad2023dynamic}. For example, the dynamic $(2-\Omega(1))$-approximate algorithms in \cite{bhattacharya2023dynamic,behnezhad2023dynamic} are inspired by the two-pass streaming algorithms (e.g.~\cite{konrad2012maximum}). Then, they use sublinear algorithms \cite{behnezhad2022time} to implement this streaming algorithm in the dynamic setting efficiently.\footnote{The dynamic $(3/2-\Omega(1))$-approximate algorithm in \cite{behnezhad2023dynamic} does not have explicit relationship to streaming algorithms. It is obtained using sublinear algorithms to improve the $(3/2)$-approximation guarantee of the tight instances of EDCS.} In contrast, it is our sublinear algorithm, not dynamic algorithm, that is inspired by the $O(1)$-pass streaming algorithm \cite{mcgregor2005finding}. Below, we explain the overview of our sublinear algorithm, which consists of two key ingredients, and then explain how our dynamic algorithm easily follows.

\subsubsection*{Ingredient I: Reduction from  $(1,\gamma n)$-Approximation to  Arbitrarily Bad Approximation.}

An initial observation is that the streaming algorithm by McGregor~\cite{mcgregor2005finding} can be viewed as the following reduction: one can compute a $(1+\gamma)$-approximate matching by making $O_{\gamma}(1)$ calls to a subroutine  that, given $S\subseteq V$, returns a $O(1)$-approximate matching of the induced subgraph $G[S]$. 

We observe that a much weaker subroutine suffices when additive approximation is allowed. Let $\largematch(S,\delta)$ be a  subroutine that, given $S\subseteq V$ and $\delta$, returns a matching $M$ in $G[S]$ such that if $\mu(G[S])\geq\delta n$, then $|M|\geq\Omega(\poly(\delta)n)$. Note that the approximation of $M$ can be arbitrarily bad depending of $\delta$. By adapting McGregor's algorithm, we show how to compute a $(1,\gamma n)$-approximate matching using only $t=O_{\gamma}(1)$ calls to 
\[
\largematch(S_{1},\delta_{1}),\dots,\largematch(S_{t},\delta_{t})
\]
where each $\delta_{i}$ is a small constant depending on $\gamma$. This algorithm, denoted by $\alg(\gamma)$, is our \emph{template} algorithm (detailed in \Cref{sec:algo:description}), which we will try to implement in the sublinear setting. 

Additionally, we observe that each vertex set $S_{i}$ can be determined in a very local manner. More precisely, a \emph{membership-query} of the form ``is a vertex $v\in S_{i}$?'' can be answered by making only $q=O_{\gamma}(1)$ \emph{matching-queries }of the form ``is a vertex $u\in V(M_{j})$? if so, return $(u,u')\in M$'' where $j<i$ and $M_{j}$ is the output of $\largematch(S_{j},\delta_{j})$ previously computed. 

However, the big challenge in the sublinear model, unlike the streaming model, is that even the weak subroutine like $\largematch(\cdot)$ is impossible.\footnote{Think of a $n\times n$ bipartite graph which consists only of a perfect matching. Using $o(n^{2})$ adjacency-matrix queries, it is not possible to out $\Omega(n)$ matching edges in this input instance. The lower bound can be extended even if we allow adjacency-list queries by adding  $\epsilon n$ dummy vertices, each of which connects to every other vertex. } Even worse, if we could not compute each matching $M_{j}$ explicitly for $j<i$, then how can we answer a membership-query whether $v\in S_{i}$? Note that known sublinear algorithms for estimating the matching size of $G[S]$ are not useful here. 

The above obstacle leads us to our second ingredient. We show that at least the oracle version of $\largematch(\cdot)$ can be implemented in the sublinear model. Later, we will explain why it is strong enough for implementing the template algorithm $\alg(\gamma)$ in the sublinear model.

\subsubsection*{Ingredient II: Large Matching Oracles on Induced Subgraphs.}

Suppose that a vertex set $A\subseteq V$ is unknown to us but a membership-query of $A$, i.e., checking if $v\in A$, can be done in $n^{1+\epsilon}$ time. 
Given access to the adjacency matrix of $G$, we show how to construct an oracle $\largematchoracle(A,\delta,\epsilon)$ with the following guarantee: 

\begin{wrapper} Using $\tilde{O}_{\delta}\left(n^{2-\epsilon}\right)$ preprocessing time, we obtain an oracle that supports matching-queries for a matching $M$ in $G[A]$ with $\tilde{O}_{\delta}\left(n^{1+g(\epsilon)}\right)$ query time where $\epsilon\le g(\epsilon)=O(\epsilon)$. If $\mu(G[A])\geq\delta n$, then $|M|=\Omega(\poly(\delta)n)$ whp.
\end{wrapper} 

The main challenge of implementing $\largematchoracle(A,\delta,\epsilon)$ in the sublinear model is that we want to find a large matching on the \emph{induced subgraph} $G[A]$. The challenge comes from possible $\Omega(n^{2})$ edges between $A$ and $V\setminus A$, and we must avoid reading these edges to get sublinear time. 
It turns out that this challenge can be overcome. 
We use the idea that appeared before in the algorithm of \cite{behnezhad2022sublinear} in a different context of estimating $(3/2-\Omega(1))$-approximation $\mu(G)$ on bipartite graphs. See the details in \Cref{sec:oracle induced}.

Given the above two ingredients, we can combine them to get our main results in the sublinear and dynamic settings, as follows.

\subsubsection*{Result I: $(1,\gamma n)$-Approximate Matching Oracles in $n^{2-\Omega_{\gamma}(1)}$ Time.}

Now, we show how to implement the template algorithm $\alg(\gamma)$ in $n^{2-\Omega_{\gamma}(1)}$ time. Let $\epsilon\in(0,1)$ be a small constant where $\lim_{\gamma\rightarrow0}\epsilon=0$. Let $\epsilon_{0}=\epsilon$ and $\epsilon_{i}=g(\epsilon_{i-1})$ for all $i\in[1,t]$ where $g$ is the function in the guarantee of Ingredient~II. So $\epsilon=\epsilon_{0}\leq\epsilon_{1}\leq\cdots\leq\epsilon_{t}$ and $\lim_{\gamma\rightarrow0}\epsilon_{t}=0$. 

We simply replace each call to $\largematch(S_{i},\delta_{i})$ with $\largematchoracle(S_{i},\delta_{i},\epsilon_{i-1})$. Now, by induction on $i\in[1,t]$, we will show that we can support membership-queries for $S_{i}$ in $\tilde{O}_{\gamma}\left(n^{1+\epsilon_{i-1}}\right)$ time and matching-queries for $M_{i}$ in $\tilde{O}_{\gamma}\left(n^{1+\epsilon_{i}}\right)$ time. Let us ignore the base case as it is trivial. For the induction step, we have the following:
\begin{enumerate}
\item To answer a membership-query for $S_{i}$, the template algorithm only needs to make $q=O_{\gamma}(1)$ matching-queries to $M_{j}$ where $j<i$. So the total query time is $q\cdot\tilde{O}_{\gamma}(n^{1+\epsilon_{i-1}})=\tilde{O}_{\gamma}(n^{1+\epsilon_{i-1}})$. 
\item To answer a matching-query for $M_{i}$, the oracle $\largematchoracle(S_{i},\delta_{i},\epsilon_{i-1})$ for the matching $M_{i}$ has query time $\tilde{O}_{\gamma}\left(n^{1+g(\epsilon_{i-1})}\right)=\tilde{O}_{\gamma}\left(n^{1+\epsilon_{i}}\right)$.
\end{enumerate}
The total preprocessing time we need for  $\largematchoracle(\cdot)$ to implement all the $t$ rounds is $\sum_{i=1}^{t}\tilde{O}_{\gamma}\left(n^{2-\epsilon_{i}}\right)=\tilde{O}_{\gamma}\left(n^{2-\epsilon}\right)=n^{2-\Omega_{\gamma}(1)}$. At the end of the last round, we can support matching-queries for the $(1,\gamma n)$-approximate matching $M$ returned by $\alg(\gamma)$ in $\tilde{O}_{\gamma}(n^{1+\epsilon_{t}})$ time, where $\lim_{\gamma\rightarrow0}\epsilon_{t}=0$.

To get a $(1,\gamma n)$-approximate estimate $\hat{\mu}$ of $\mu(g)$, we  sample $\tilde{O}(1/\gamma^2)$ vertices and check if they are matched under $M$. Whp, this is a $(1,\Theta(\gamma)n)$-approximation of $\mu(G)$ because $M$ is $(1,\gamma n)$-approximate.

\subsubsection*{Result II: Dynamic $(1+\gamma)$-Approximate Matching Size.}

Our dynamic matching size algorithm now follows from standard techniques. Using the well-known vertex reduction technique (see, for example, Corollary 4.9 of \cite{kiss2022improving}), we can assume that $\mu(G)\ge\gamma n$ at all times. We work in \emph{phases}, where each phase lasts for $\gamma^{2}n$ updates. At the start of each phase, we invoke the sublinear algorithm from Result I above, to obtain a $(1,\gamma^{2}n)$-approximate estimate $\hat{\mu}$ of $\mu(G)$, in $n^{2-\Omega_{\gamma}(1)}$ time. Since $\mu(G)\geq\gamma n$ and since the phase lasts for only $\gamma^{2}n$ updates, this $\hat{\mu}$ continues to remain a purely multiplicative $(1+\Theta(\gamma))$-approximate estimate of $\mu(G)$ throughout the duration of the phase. This leads to an amortized update time of $n^{2-\Omega_{\gamma}(1)}/(\gamma^{2}n)=n^{1-\Omega_{\gamma}(1)}$. In Section~\ref{sec:dynamic}, we show how to extend this approach to prove Theorem~\ref{thm:main dynamic}.

%% file: 3-prelim.tex
\section{Notations and Preliminaries}
\label{sec:prelim}


Unless speficied otherwise, the input graph $G = (V, E)$ will have $n$ nodes and $m$ edges. A {\em matching} $M \subseteq E$ is a subset of edges that do not share any common endpoint. We use the symbol $\mu(G)$ to denote the size of a maximum matching in $G$. We say that a pah $p = (v_0, v_1, \ldots, v_i)$ is an {\em alternating path} in $G$ w.r.t.~a matching $M \subseteq E$ iff $(v_{j}, v_{j+1}) \in E$ for all $j \in [0, i-1]$ and the edges in the path $p$ alternate between being in $M$ and in $E \setminus M$. We say that $p$ is an {\em augmenting path} in $G$ w.r.t.~$M$ iff $p$ is an alternating path whose first and the last edges are both unmatched in $M$. The {\em length} of a path is the number of edges in it. We let $V(M)$ denote the set of matched nodes in  a matching $M \subseteq E$. Consider any node $v \in V(M)$ and suppose that $(u, v) \in M$. Then we say that $u$ is the {\em mate} of $v$ in $M$. Given a subset of nodes $S \subseteq V$, $G[S]$ denotes the subgraph of $G$ induced by $S$. Given any graph $G'$, the symbol $E(G')$ denotes the set of edges in $G'$. 

Throughout the paper, the symbol $\Theta_{k, \gamma}(1)$ will denote any positive constant that depends only on $k$ and $\gamma$ (where $k$ and $\gamma$ are constant parameters  whose values will be chosen later on). We analogously use the notation $\Theta_k(1)$ to denote a constant that depends only on $k$. Finally, the symbol $\tilde{O}(.)$ will be used to hide any $\polylog (n)$ factors.

\paragraph{Oracles.}
We have the adjacency matrix access to the input graph $G$. Each query takes $O(1)$ time. We do \emph{not} have the adjacency list access to the input graph.

For any vertex set $A\subset V$, an \emph{$A$-membership} oracle $\mem_{A}:V\rightarrow\{0,1\}$ indicates whether $v\in A$ for any $v\in V$. That is, we have 
\[
\mem_{A}(v)=\boldsymbol{1}\{v\in A\}.
\]
A \emph{matching oracle $\match_M:V\rightarrow\binom{V}{2}\cup\{\bot\}$ for a matching} $M$ is an oracle that, given a vertex $v\in V$, returns 
\[
\match_M(v)=\begin{cases}
(v,v') & (v,v')\in M\\
\bot & v\notin V(M).
\end{cases}
\]
Similarly, a {\em mate oracle $\mate_M : V \rightarrow V \cup \{ \perp\}$ for a matching} $M$ is an oracle that, given a vertex $v \in V$, returns 
\[
\mate_M(v)=\begin{cases}
v' & v \in V(M) \text{ and } (v,v')\in M\\
\bot & v\notin V(M).
\end{cases}
\]

\paragraph{Concentration Bounds.} We  need standard concentration bounds as follows.

\begin{prop}
[Hoeffding bound]\label{prop:hoeffding}Let $X_{1},\dots,X_{n}$ be independent random variables such that $a\le X_{i}\le b$. Let $X=\sum_{i=1}^{n}X_{i}$. For any $t>0$, 
\[
\Pr[|X-\E[X]|\ge t]\le2\exp(-\frac{2t^{2}}{n(b-a)^{2}}).
\]
\end{prop}

\begin{prop}
[Chernoff bound]\label{prop:chernoff}Let $X_{1},\dots,X_{n}$ be independent $\{0,1\}$-random variables. Let $X=\sum_{i=1}^{n}X_{i}$ where $\E[X]\le\overline{\mu}$ For any $t>0$ where $t\le\overline{\mu}$, 
\[
\Pr[|X-\E[X]|\ge t]\le2\exp(-\frac{t^{2}}{3\overline{\mu}}).
\]
\end{prop}
Chernoff bound can be much stronger than Hoeffding bound when $\E[X]$ has small upper bound. For example, if we applied \Cref{prop:hoeffding} to the setting for \Cref{prop:chernoff}, we would only get that the bound of $2\exp(-\frac{2t^{2}}{3n})$ which is much weaker than $2\exp(-\frac{t^{2}}{3\overline{\mu}})$ when $\overline{\mu}\ll n$.


%% file: 4-matching-oracle.tex
\section{Matching Oracles of Induced Subgraphs}

\label{sec:oracle induced}

In this section, we present the key subroutine of this paper. The goal is to construct a matching oracle for an induced subgraph $G[A]$ but $A$ is unknown to us; we only have access to an $A$-membership oracle $\mem_{A}$.

\begin{thm}
\label{thm:matching oracle}Let $G=(V,E)$ be a graph, $A\subseteq V$ be a vertex set. Suppose that we have access to adjacency matrix of $G$ and an $A$-membership oracle $\mem_{A}$ with $t_{A}$ query time. We are given as input $\epsilon>0$ and $\sizeIn>0$.

We can preprocess $G$ in $\Otil((t_{A}+n)(n^{1-\epsilon}+n^{4\epsilon})/\poly(\sizeIn))$   time and either return $\bot$ or construct a matching oracle $\match_{M}(\cdot)$ for a matching $M\subset G[A]$ of size at least $\sizeOut n$ where $\sizeOut=\sizeIn^{5}/{10^8}$ that has $\Otil((t_{A}+n)n^{4\epsilon}/\poly(\sizeIn))$ 
worst-case query time. If $\mu(G[A])\ge\sizeIn n$, then $\bot$ is not returned. The guarantee holds with high probability.
\end{thm}\thatchaphol{I changed $2000$ to $10^8$.}

The very important property of \Cref{thm:matching oracle} is that it makes $n^{1-\epsilon}$ oracle calls to $\mem_{A}$ during preprocessing and only $n^{O(\epsilon)}$ calls to $\mem_{A}$ on each query. The rest of this section is devoted for proving \Cref{thm:matching oracle}. 

To prove \Cref{thm:matching oracle}, we adapt the technique used inside the algorithm by Behnazhad et al.~\cite{behnezhad2022sublinear} for $(3/2-\Omega(1))$-approximating $\mu(G)$ on bipartite graphs. We observe that the idea there has reach beyond $(3/2-\Omega(1))$-approximation algorithms. The abstraction of that idea leads us to \Cref{thm:matching oracle}, the crucial subroutine for later parts of our paper.

This section is organized as follows. In \Cref{sec:induced lowdeg}, we show a weaker version of \Cref{thm:matching oracle} that works well on low degree graphs. We will use this weaker version in the preprocessing step, described in \Cref{sec:induced prep}. Then, we complete the query algorithm in \Cref{sec:induced query}.

\subsection{Oracles on Low Degree Graphs}

\label{sec:induced lowdeg}

Here, we show a similar result as \Cref{thm:matching oracle}, but it is efficient only when the maximum degree $\Delta$ is small. In particular, the query algorithm makes $n^{O(\epsilon)}$ calls to $\mem_{A}$ only when $\Delta=n^{O(\epsilon)}$. 
\begin{lem}
\label{lem:lowdeg oracle}Let $G=(V,E)$ be a graph with maximum degree $\Delta$ where $\Delta$ is known and $A\subseteq V$ be a vertex set. Suppose that we have access to adjacency matrix of $G$ and an $A$-membership oracle $\mem_{A}$ with $t_{A}$ query time. We can construct in $\Otil((t_{A}\Delta+n+t_{A}/\epsilon)\Delta/\epsilon^{2})$ time a matching oracle $\match_{M}^{\lowdeg}(\cdot)$ for a $(2,\epsilon n)$-approximate matching $M$ in $G[A]$ that has $\Otil(t_{A}\Delta+n+t_{A}/\epsilon)\Delta/\epsilon)$ worst-case query time with high probability.
\end{lem}

The proof of \Cref{lem:lowdeg oracle} is based on the the following $(2,\epsilon n)$-approximate matching oracle given access to adjacency list.
\begin{lem}
\label{thm:simple LCA}Given the adjacency lists of a graph $G=(V,E)$ with average degree $d$ and a parameter $\dbar\ge d$, we can in $\Otil(\dbar/\epsilon^{2})$ time construct a matching oracle $\match_{M}(\cdot)$ for a $(2,\epsilon n)$-approximate matching $M$ in $G$ with $\Otil(\dbar/\epsilon)$ worst-case query time with high probability.
\end{lem}

\Cref{thm:simple LCA} is proved by combining an improved analysis of randomized greedy maximal matching of Behnezhad \cite{behnezhad2022time} into a framework for constructing an LCA by \cite{levi2015local}. We do not claim any novel contribution here and defer the proof to \Cref{sec:simple LCA}.

Now, to prove \Cref{lem:lowdeg oracle}, we need to strengthen \Cref{thm:simple LCA} in two ways. First, it must work with the adjacency matrix, not the adjacency lists. Second, it must return a large matching of an induced subgraph $G[A]$, not that of $G$. However, this can be done using a simple simulation.
\begin{proof}
[Proof of \Cref{lem:lowdeg oracle}]Let ${\cal A}$ denote the algorithm of \Cref{thm:simple LCA}. We simulate ${\cal A}$ on $G[A]$ with parameter $\dbar\gets\Delta$ as follows.

Whenever ${\cal A}$ needs to sample a vertex, we sample $O(\log(n)/\epsilon)$ vertices in $G$ and call $\mem_{A}$ on each of them. If one of them is in $A$, then we get a random vertex in $G[A]$. If none of them is in $A$, then w.h.p. $|A|\le\epsilon n$. If this ever happens, even an empty matching is a $(2,\epsilon n)$-approximate matching in $G[A]$, and the problem becomes trivial. 

Whenever ${\cal A}$ needs to make queries to the adjacency list of any vertex $v$, we can construct the whole adjacency list of $v$ in $G[A]$ by first making $n$ adjacency matrix queries to learn all neighbors of $v$ in $G$ and then makes $\deg(v)\le\Delta$ oracles calls to $\mem_{A}$ to know which neighbors are in $G[A]$. This takes $O(t_{A}\Delta+n)$ time. Every other computation can be simulated without the overhead. 

Therefore, each step of ${\cal A}$ can be simulated with an extra $(t_{A}\cdot O(\log(n)/\epsilon)+t_{A}\Delta+n)$ factor. 
\end{proof}

\subsection{Preprocessing}

\label{sec:induced prep}

We describe the preprocessing algorithm in \Cref{alg:prep} with the guarantees summarized in the lemma below.
\begin{lem}
\label{lem:prep}In $\Otil((t_{A}+n)(n^{1-\epsilon}+n^{4\epsilon})/\poly(\sizeIn))$ time, \Cref{alg:prep} outputs either $\bot$ (indicating an error) or \emph{the remaining} set $V'\subseteq V$ of vertices together with an explicit matching $M'\subseteq G[V']$ that satisfies one of the following:
\begin{enumerate}
\item \label{enu:explicit}$|M'[A]|\ge2\sizeOut n$, or
\item \label{enu:implicit}$\mu(G[A\cap V'\setminus V(M')])\ge4\sizeOut n$ and $G[V'\setminus V(M')]$ has maximum degree at most $n^{2\epsilon}$.
\end{enumerate}
The algorithm also reports which properties above $M'$ satisfies. If $\mu(G[A])\ge\sizeIn n$, then $\bot$ is not returned with high probability.
\end{lem}

In \Cref{alg:prep}, the remaining set $V'$ is initialized as $V$ and only shrinks. For convenience, we let $A':=A\cap V'$ and $D':=D\cap V'$ denote the \emph{remaining} alive and dead vertices.\sayan{Where are we defining $A$ and $D$ as ``alive'' and ``dead'' vertices? Where are we initializing $D$?} 

\begin{algorithm}
$p=100n^{2-2\epsilon}\log n$, $k=n^{\epsilon}$, $\eta=\sizeIn^{2}\log(n)/10$, $T=100/\sizeIn^{2}$, $\sizeOut=\sizeIn^{3}/10^6T=\sizeIn^{5}/10^8$.

$r_{1}=r_{2}=\frac{1000}{\sizeOut^{2}}\log n=\Theta(\frac{\log n}{\sizeIn^{10}})$, $r_{3}=1000\sizeIn\frac{n}{k}\log n$.

$V'\gets V$.

\textbf{Repeat the following for $T$ times:}
\begin{enumerate}
\item Sample $kp$ distinct pairs of vertices from $V'.$ Partition the sampled pairs into $(P^{1},\dots,P^{k})$ where each $P^{i}$ is an ordered list containing $p$ pairs of vertices. 
\item For $i\in[k]$
\begin{enumerate}
\item Let $E^{i}=\{(u,v)\in P^{i}\mid(u,v)\in G[V'] \}$ be an ordered sublist of $P^{i}$ containing only pairs which are edges of $G[V']$.
\item Let $M^{i}$ be the greedy maximal matching when scanning $E^{i}$ in order. 

\textbf{\textbackslash{}\textbackslash{}Case \ref{enu:explicit}:}
\item Sample $r_{1}$ edges from $M^{i}$. 
\item Let $X$ count the sampled edges that are in $G[A]$ (using the oracle $\mem_{A}$)
\item Set $\tilde{\mu}_{1}=\frac{|M^{i}|X}{r_{1}}-\frac{\sizeOut n}{2}$. 
\item \label{line:terminate case explicit}If $\tilde{\mu}_{1}\ge2\sizeOut n$, then set $M'\gets M^{i}$ and \textbf{report} that $M'$ satisfies Case \ref{enu:explicit}.

\textbf{\textbackslash{}\textbackslash{}Case \ref{enu:implicit}:}
\item \label{enu:define lowdeg matching}Let $\Mhat^{i}$ be a $(2,\sizeOut n)$-approximate matching in $G[A'\setminus V(M^{i})]$ that the matching oracle $\match_{\Mhat^{i}}^{\lowdeg}(\cdot)$ from \Cref{lem:lowdeg oracle} respects, given graph $G[V'\setminus V(M^{i})]$ with vertex set $A'\setminus V(M^{i})$ as input.
\item Sample $r_{2}$ vertices from $V'\setminus V(M^{i})$. 
\item Let $Y$ count the sampled vertices that are matched in $\Mhat^{i}$ (using the oracle $\match_{\Mhat^{i}}^{\lowdeg}(\cdot)$).
\item Set $\tilde{\mu}_{2}=\frac{|V'\setminus V(M^{i})|Y}{2r_{2}}-\frac{\sizeOut n}{2}$.
\item \label{line:terminate case implicit}If $\tilde{\mu}_{2}\ge4\sizeOut n$, then set $M'\gets M^{i}$ and \textbf{report} that $M'$ satisfies Case \ref{enu:implicit}.
\end{enumerate}
\item \label{line:Case3}Let $A'_{\sp}\subseteq A'$ be obtained by sampling $r_{3}$ vertices from $A'$.
\item Let $\Gbar=(V',\cup_{i=1}^{k}M^{i})$.
\item Let $C=\{v\in V'\mid N_{\Gbar}(v,A'_{\sp})\ge\eta\}$, i.e., $C$ contains remaining vertices that have at least $\eta$ neighbors from $A'_{\sp}$ in $\Gbar$. 
\item Set $V'\gets V'\setminus C$.
\end{enumerate}
\textbf{Return $\bot$ (Error).}

\caption{Preprocess $G$.\label{alg:prep}}

\end{algorithm}\sayan{In step 3 of Algorithm 1, how do we sample vertices from $A'$ when we don't know what $A'$ is?}

\subsubsection{Correctness }

In this part, we prove the correctness of \Cref{alg:prep} assuming that it does not return $\bot$. We first show that $\tilde{\mu}_{1}$ and $\tilde{\mu}_{2}$ are good approximation of $M^{i}[A]$ and $\Mhat^{i}$ base on basic there definition and Hoeffding's bound.
\begin{lem}
\label{lem:size explicit}For every $i$, we have $|M^{i}[A]|-\sizeOut n\le\tilde{\mu}_{1}\le|M^{i}[A]|$ w.h.p.
\end{lem}

\begin{proof}
The probability that a random edge from $M^{i}$ is in $G[A]$ is $\frac{|M^{i}[A]|}{|M^{i}|}$. So $\E[X]=r_{1}\frac{|M^{i}[A]|}{|M^{i}|}$ and $|M^{i}[A]|=\frac{|M^{i}|\E[X]}{r_{1}}$. By definition of $\tilde{\mu}_{1}=\frac{|M^{i}|X}{r_{1}}-\frac{\sizeOut n}{2}$ and by Hoeffding bound \Cref{prop:hoeffding}, we have
\begin{align*}
\Pr[\tilde{\mu}_{1}<|M^{i}[A]|-\sizeOut n\text{ or }\tilde{\mu}_{1}>|M^{i}[A]|] & =\Pr[\frac{|M^{i}|\cdot(X-\E[X])}{r_{1}}>\frac{\sizeOut n}{2}]\\
 & \le2\exp(-\frac{2(\frac{\sizeOut n}{2})^{2}}{r_{1}(\frac{n}{r_{1}})^{2}})=2\exp(-\sizeOut^{2}r_{1}/2)\\
 & \le1/n^{10}.
\end{align*}
\end{proof}
\begin{lem}
\label{lem:size implicit}For every $i$, we have $|\Mhat^{i}|-\sizeOut n\le\tilde{\mu}_{2}\le|\Mhat^{i}|$ w.h.p.
\end{lem}

\begin{proof}
The probability that a random vertex from $V'\setminus V(M^{i})$ is in $V(\Mhat^{i})$ is $\frac{2|\Mhat^{i}|}{|V'\setminus V(M^{i})|}$. So $\E[Y]=r_{2}\frac{2|\Mhat^{i}|}{|V'\setminus V(M^{i})|}$ and $|\Mhat^{i}|=\frac{|V'\setminus V(M^{i})|\E[Y]}{2r_{2}}$. By definition of $\tilde{\mu}_{2}=\frac{|V'\setminus V(M^{i})|Y}{2r_{2}}-\frac{\sizeOut n}{2}$ and by Hoeffding bound \Cref{prop:hoeffding}, we have
\begin{align*}
\Pr[\tilde{\mu}_{2}<|\Mhat^{i}|-\sizeOut n\text{ or }\tilde{\mu}_{2}>|\Mhat^{i}|] & =\Pr[\frac{|V'\setminus V(M^{i})|\cdot(Y-\E[Y])}{2r_{2}}>\frac{\sizeOut n}{2}]\\
 & \le2\exp(-\frac{2(\frac{\sizeOut n}{2})^{2}}{r_{2}(\frac{n}{2r_{2}})^{2}})=2\exp(-2\sizeOut^{2}r_{2})\\
 & \le1/n^{10}.
\end{align*}
\end{proof}
Next, we show the ``sparsification'' property of randomized greedy maximal matching $M^{i}$. That is, $G[V'\setminus V(M^{i})]$ has low degree. The idea is that any high-degree vertex $v$ in $G[V'\setminus V(M^{i})]$ should not exist because it should have been matched by $M^{i}$ via one of the sampled edges. The proof is similar to Lemma 3.1 of \cite{blelloch2012greedy}, which considers this sparsification property of the randomized greedy maximal independent set, instead of maximal matching.
\begin{lem}
\label{lem:sparsification}For every $i$, the maximum degree of $G[V'\setminus V(M^{i})]$ is at most $n^{2\epsilon}$ w.h.p. 
\end{lem}

\begin{proof}
Let us describe an equivalent way to construct $M^{i}$. Initialize $M^{i}=\emptyset$ and then sample $p$ pairs of vertices in $V'$. For each sampled pair $(u,v)$, if $(u,v)\in G[V']$ and both $u$ and $v$ are not matched by $M^{i}$, then we add $(u,v)$ into $M^{i}$. At the end of this process, we will show that, for any vertex $v\in V'$, the degree of $v$ in $G[V'\setminus V(M^{i})]$ is at most $n^{2\epsilon}$ w.h.p. (we use the convention that if $v\notin V'\setminus V(M^{i})$, then the degree $v$ is $0$.) 

For $t\in[1,p]$, let $M_{t}^{i}$ denote the matching $M^{i}$ after we sampled the $t$-th pair. For convenience, we denote $\deg^{t}(v)=\deg_{G[V'\setminus V(M_{t}^{i})]}(v)$ as the \emph{degree of $v$ at time $t$.} We want to show that $\Pr[\deg^{p}(v)>n^{2\epsilon}]\le1/n^{10}$ for any $v\in V'$. 

Observe that if $\deg^{p}(v)>n^{2\epsilon}$, then $\deg^{t}(v)>n^{2\epsilon}$ for all $t\le p$. Now, given that $\deg^{t-1}(v)>n^{2\epsilon}$, the probability that $v$ remained at unmatched after time $t$ is 
\[
1-\frac{\deg^{t-1}(v)}{\binom{|V'|}{2}}\le1-\frac{n^{2\epsilon}}{n^{2}}.
\]
In particular, the probability that $\deg^{t}(v)>n^{\epsilon}$ is at most $1-\frac{n^{2\epsilon}}{n^{2}}$. This implies that the probability that $\deg^{p}(v)>n^{2\epsilon}$ is at most 
\[
(1-\frac{n^{2\epsilon}}{n^{2}})^{p}\le\frac{1}{n^{10}}
\]
because $p=100n^{2-2\epsilon}\log n$.
\end{proof}
From the above lemmas, we can conclude the correctness of the algorithm. 
\begin{cor}
\label{cor:prep correct}If \Cref{alg:prep} returns a matching $M'$, then $M'$ satisfies the guarantees from \Cref{lem:prep} w.h.p.
\end{cor}

\begin{proof}
If $M'$ is returned under Case~\ref{enu:explicit}. Then, by \Cref{lem:size explicit}, we have $|M'[A]|\ge\tilde{\mu}_{1}\ge2\sizeOut n$ w.h.p. Otherwise, in Case~\ref{enu:implicit}, we have $\mu(G[A\cap V'\setminus V(M')])\ge|M'|\ge\tilde{\mu}_{2}\ge4\sizeOut n$ by \Cref{lem:size implicit} and also maximum degree of $G[V'\setminus V(M')]$ is at most $n^{2\epsilon}$ by \Cref{lem:sparsification} w.h.p.
\end{proof}

\subsubsection{Termination without Error}

In this part, we show that if $\mu(G[A])\ge\sizeIn n$, then \Cref{alg:prep} does not return $\bot$ w.h.p. For any graph $G$ and $U_{1},U_{2}\subseteq V(G)$, we let $G[U_{1},U_{2}]$ contains all edges of $G$ whose one endpoint is in $U_{1}$ and another in $U_{2}$. Note that the induced subgraph $G[U_{1}]=G[U_{1},U_{1}]$. 

Our high-level plan is that we will show that $D'$ decreases its size by $\Theta(\sizeIn^{2}n)$ in each iteration in the repeat loop. So $D'$ must become very small after $T=\Theta(1/\sizeIn^{2})$ iterations. But when this happens, we can show that either Case~\ref{enu:explicit} or Case~\ref{enu:implicit} must happen and so the algorithm must terminate without error. 

To carry out the above plan, we need a helper lemma (\Cref{lem:large matching remain}) which states that, even if $V'$ keeps shrinking, the maximum matching in $G[V']$ remains large, $\mu(G[A'])\ge\sizeIn n/2$. We will need this fact throughout the whole argument. 

The high-level argument goes as follows. If the algorithm does not return $M'$, then $M^{i}[A']$ is very small for all $i$ and so $\Gbar[A']$ contains few edges. It follows that the set $C$ of removed vertices contains only few vertices in $A'$ because each vertex $v\in C\cap A'$ has high degree of at least $\eta$ in $\Gbar[A']$. Thus, we remove only few vertices from $A'$ and so the size of $\mu(G[A'])$ cannot decrease too much. The formal argument below goes through the set $A'_{\sp}$ and the above paragraph gives the right intuition. 
\begin{lem}
\label{lem:large matching remain}Suppose $\mu(G[A])\ge\sizeIn n$. For $\tau\in[0,T]$, at the end of the $\tau$-th iteration of the repeat loop in \Cref{alg:prep}, we have $\mu(G[A'])\ge(1-\frac{\tau}{2T})\sizeIn n\ge\sizeIn n/2$ w.h.p., if the algorithm does not terminate yet.
\end{lem}

\begin{proof}
We prove by induction on $\tau$. For $\tau=0$ (i.e. the beginning the algorithm), the claim holds as $\mu(G[A])\ge\sizeIn n$. Next, we consider $\tau\ge1$. By induction hypothesis, at the beginning of the $\tau$-th iteration, we have $\mu(G[A'])\ge(1-\frac{\tau-1}{2T})\sizeIn n\ge\sizeIn n/2$. 

At the end of the $\tau$-th iteration, since \Cref{alg:prep} did not terminate at Step \ref{line:terminate case explicit}, by \Cref{lem:size explicit}, we have, w.h.p., $|M^{i}[A']|\le2\sizeOut n+\sizeOut n\le3\sizeOut n$ for all $i$. So we have $|E(\Gbar[A'])|\le3\sizeOut nk$ and then the average degree of vertices in $\Gbar[A']$ is at most
\[
\frac{2|E(\Gbar[A'])|}{|A'|}\le\frac{6\sizeOut k}{\sizeIn}
\]
because $|A'|\ge2\mu(G[A'])\ge\sizeIn n$. 

Recall that $A'_{\sp}$ is obtained by sampling $r_{3}$ vertices from $A'$. So 
\[
\E[\vol_{\Gbar[A']}(A'_{\sp})]\le\frac{6\sizeOut kr_{3}}{\sizeIn}\le\frac{\sizeIn^{3}n\log n}{200T}
\]
because $\sizeOut=\sizeIn^{3}/10^{6}T$ and $\frac{kr_{3}}{\sizeIn}=1000n\log n$. Furthermore, this bound is concentrated. Indeed, since $\vol_{\Gbar[A']}(A'_{\sp})$ is a sum of $r_{3}$ independent random variable whose range is $k$ (since the maximum degree in $\Gbar$ and $\Gbar[A']$ is $k$), by Hoeffding bound \Cref{prop:hoeffding}, we have
\[
\Pr[\vol_{\Gbar[A']}(A'_{\sp})-\E[\vol_{\Gbar[A']}(A'_{\sp})]>\frac{\sizeIn^{3}n\log n}{200T}]\le2\exp(\frac{-2(\frac{\sizeIn^{3}n\log n}{200T})^{2}}{r_{3}k^{2}})\ll1/n^{10}.
\]
So $\vol_{\Gbar[A']}(A'_{\sp})\le\frac{\sizeIn^{3}n\log n}{100T}$ w.h.p. Since every vertex $v\in C$ is adjacent to at least $\eta$ vertices in $A'_{\sp}$ in $\Gbar$, we have $|C\cap A'|\eta\le\vol_{\Gbar[A']}(A'_{\sp})$. Therefore,
\[
|C\cap A'|\le\frac{\sizeIn^{3}n\log n}{100T\eta}\le\frac{\sizeIn n}{2T}
\]
because $\eta=\sizeIn^{2}\log(n)/10$. This means that we remove at most $\frac{\sizeIn n}{2T}$ vertices from $A'$ at the end of the $\tau$-th iteration. So the size of maximum matching in $G[A']$ may decrease by at most $\frac{\sizeIn n}{2T}$. Thus, $\mu(G[A'])\ge(1-\frac{\tau}{2T})\sizeIn n$ which completes the induction.
\end{proof}
Given \Cref{lem:large matching remain}, we will use the following lemma to argue that if \Cref{alg:prep} does not return $M^{i}$, then then $M^{i}$ must match many vertices between $A'$ and $D'$. 
\begin{lem}
\label{lem:match A and D}Suppose $\mu(G[A'])\ge\sizeIn n/2$. For any matching $M$ in $G[V']$, if $|M[A']|<3\sizeOut n$ and $\mu(G[A'\setminus V(M)])<16\sizeOut n$, then $M[A',D']\ge\sizeIn n/3$.
\end{lem}

\begin{proof}
Let $M^{*}$ be the maximum matching in $G[A']$ of size at least $\sizeIn n/2$. We partition edges in $M^{*}$ into two parts: $M_{0}^{*}$ and $M_{1}^{*}$. For each $(u,v)\in M^{*}$, we add $(u,v)$ into $M_{0}^{*}$ if both $u,v\notin V(M)$. Otherwise, either $u$ or $v$ are matched by $M$ and we add $(u,v)$ into $M_{1}^{*}$. Note that $M_{0}^{*}$ is a matching in $G[A'\setminus V(M)]$. So $|M_{0}^{*}|<16\sizeOut n$ and $|M_{1}^{*}|\ge\sizeIn n/2-16\sizeOut n$. 

Observe that $|V(M_{1}^{*})|\le|M[A',D']|+2|M[A']|$ because we can charge vertices of $V(M_{1}^{*})$ to either matched edges of $M[A',D']$ or $M[A']$ such that each matched edge in $M[A',D']$ is charged once and each match edges in $M[A']$ is charged at most twice. Since $|M[A']|<3\sizeOut n$, we have $|M[A',D']|\ge\sizeIn n/2-16\sizeOut n-2\cdot3\sizeOut n\ge\sizeIn n/3$.
\end{proof}
\Cref{lem:match A and D} also says that once $D'$ become small enough, \Cref{alg:prep} will not err w.h.p. 
\begin{cor}
\label{cor:condition terminate}If $\mu(G[A'])\ge\sizeIn n/2$ and $|D'|<\sizeIn n/3$, then \Cref{alg:prep} will return a matching $M'$ w.h.p. 
\end{cor}

\begin{proof}
For any $i$, note that $|M^{i}[A',D']|<|D'|<\sizeIn n/3$. So by the contrapositive of \Cref{lem:match A and D}, we have that either $|M^{i}[A']|\ge3\sizeOut n$ or $\mu(G[A'\setminus V(M^{i})])\ge10\sizeOut n$. If $|M^{i}[A']|\ge3\sizeOut n$, then $\tilde{\mu}_{1}\ge2\sizeOut n$ w.h.p.~by \Cref{lem:size explicit}. If $\mu(G[A'\setminus V(M^{i})])\ge10\sizeOut n$, then $|\Mhat_{i}|\ge\frac{16\sizeOut n}{2}-\sizeOut n\ge7\sizeOut n$ because $\Mhat_{i}$ is $(2,\sizeOut n)$-approximate matching. So $\tilde{\mu}_{2}\ge6\sizeOut$ w.h.p.~by \Cref{lem:size implicit}. In either cases, so \Cref{alg:prep} must return a matching $M'$ at Line~\ref{line:terminate case explicit} or Line~\ref{line:terminate case implicit}.
\end{proof}
Now, we are ready to show that $D'$ must shrink significantly after each iteration of the repeat loop, which means that there cannot be too many iterations before the algorithm terminate by \Cref{cor:condition terminate}.
\begin{lem}
\label{lem:D shrinks}Suppose$\mu(G[A'])\ge\sizeIn n/2$. If \Cref{alg:prep} does not terminate until $C$ is computed, then $|D'\cap C|\ge\frac{\sizeIn^{2}}{100}n$. 
\end{lem}

There are two main claims in the proof of \Cref{lem:D shrinks}. We suggest the reader to skip the proofs of these claims and see how they are used to prove \Cref{lem:D shrinks} first. 
\begin{claim}
\label{claim:lower bound edges}$|E(\Gbar[A'_{\sp},D'])|\ge100\sizeIn^{2}n\log n$ w.h.p. 
\end{claim}

\begin{proof}
At the end of the $\tau$-th iteration, since \Cref{alg:prep} did not terminate at Step~\ref{line:terminate case explicit} nor Step~\ref{line:terminate case implicit}, we have, w.h.p., that $M[A']<3\sizeOut n$ by \Cref{lem:size explicit} and $|\Mhat^{i}|<5\sizeOut n$ by \Cref{lem:size implicit}. Since $\Mhat^{i}$ is a $(2,\sizeOut n)$-approximate matching in $G[A'\setminus V(M)]$, we have $\mu(G[A'\setminus V(M)])<16\sizeOut n$. By \Cref{lem:match A and D}, we have $|M^{i}[A',D']|\ge\sizeIn n/3$.

Observe that $|E(\Gbar[A',D'])|=\sum_{i}|M^{i}[A',D']|\ge\sizeIn nk/3$ because all $M^{i}$ are\textbf{ }mutually disjoint. Note that $\Gbar[A',D']$ is a bipartite graph. So the average degree of vertices in $A'$ in $\Gbar[A',D']$ is 
\[
\frac{|E(\Gbar[A',D'])|}{|A'|}\ge\sizeIn k/3
\]
and we have that 
\[
\E[|E(\Gbar[A'_{\sp},D'])|]\ge\sizeIn kr_{3}/3\ge200\sizeIn^{2}n\log n.
\]
because $r_{3}=1000\sizeIn\frac{n}{k}\log n.$ Furthermore, this is concentrated. Indeed, since $|E(\Gbar[A'_{\sp},D'])|$ is a sum of $r_{3}$ independent random variable whose range is $k$ (since the maximum degree in $\Gbar$ and $\Gbar[A']$ is $k$), by Hoeffding bound \Cref{prop:hoeffding}, we have
\[
\Pr[\left||E(\Gbar[A'_{\sp},D'])|-\E[|E(\Gbar[A'_{\sp},D'])]|]\right|>100\sizeIn^{2}n\log n]\le2\exp(\frac{-2(100\sizeIn^{2}n\log n)^{2}}{r_{3}k^{2}})\ll1/n^{10}.
\]
So $|E(\Gbar[A'_{\sp},D'])|\ge100\sizeIn^{2}n\log n$ w.h.p.
\end{proof}
\begin{claim}
\label{claim:upper bound degree}For each $v\in D'$, the number of neighbor of $v$ from $A'_{\sp}$ in $\Gbar$ is at most $2000\log n$ w.h.p. That is, $|N_{\Gbar}(v,A'_{\sp})|\le2000\log n$. 
\end{claim}

\begin{proof}
We have 
\begin{align*}
\E[N_{\Gbar}(v,A'_{\sp})] & =\sum_{u\in N_{\Gbar}(v,A')}\Pr[u\in A'_{\sp}]\\
 & \le k\cdot\frac{r_{3}}{|A'|}\le1000\log n
\end{align*}
because $r_{3}k=1000\sizeIn n\log n$ and $|A'|\ge\sizeIn n$ since we assume $\mu(G[A'])\ge\sizeIn n/2$. Moreover, applying by Chernoff bound \Cref{prop:chernoff} where $t=1000\log n$ and $\overline{\mu}=1000\log n$\footnote{Note that Hoeffding bound \Cref{prop:hoeffding} is not strong enough here.}, we have
\[
\Pr[|N_{\Gbar}(v,A'_{\sp})-\E[N_{\Gbar}(v,A'_{\sp})|>1000\log n]\le2\exp(-\frac{(1000\log n)^{2}}{3\cdot1000\log n})\ll1/n^{10}.
\]
So $|N_{\Gbar}(v,A'_{\sp})|\le2000\log n$ w.h.p. 
\end{proof}
Now, let us prove \Cref{lem:D shrinks} using the above claims. 

\paragraph{Proof of \Cref{lem:D shrinks}.}
Observe that 
\begin{align*}
|E(\Gbar[A'_{\sp},D'])| & =\sum_{v\in D'}|N_{\Gbar}(v,A'_{\sp})|\le|D'\cap C|\cdot2000\log n+|D'\setminus C|\eta
\end{align*}
where the inequality holds w.h.p.~by \Cref{claim:upper bound degree}. Since $|D'\setminus C|\le n$ and $\eta=\sizeIn^{2}\log(n)/10$, we have by \Cref{claim:lower bound edges} that
\[
100\sizeIn^{2}n\log n\le|D'\cap C|\cdot2000\log n+n\sizeIn^{2}\log(n)/10
\]
and so $|D'\cap C|\ge\frac{\sizeIn^{2}}{100}n$ as desired. \hfill$\square$

Finally, we give the conclusion of this part. 
\begin{cor}
\label{cor:conclude terminate}If $\mu(G[A])\ge\sizeIn n$, then \Cref{alg:prep} does not return $\bot$ w.h.p.
\end{cor}

\begin{proof}
First, $\mu(G[A])\ge\sizeIn n$ implies that $\mu(G[A'])\ge\sizeIn n/2$ w.h.p.~by \Cref{lem:large matching remain}. So, by \Cref{lem:D shrinks} $D'$ decreases its size by $\sizeIn^{2}n/100$ in each iteration in the repeat loop. Hence, we have that $|D'|\le\sizeIn n/3$ before $T=100/\sizeIn^{2}$ iterations. Therefore, there is an iteration $\tau\in[1,T]$ where \Cref{alg:prep} will return a matching $M'$ w.h.p. by \Cref{cor:condition terminate}.
\end{proof}

\subsubsection{Preprocessing Time}

Consider the $(2,\sizeOut n)$-approximate matching oracle $\match^{\lowdeg}(\cdot)$ in Line~\ref{enu:define lowdeg matching}, which is given graph $G[V'\setminus V(M^{i})]$ and vertex set $A'\setminus V(M^{i})$ as input. 

By \Cref{lem:sparsification}, we can assume w.h.p. that $G[V'\setminus V(M^{i})]$ has degree at most $n^{2\epsilon}$. \Cref{lem:lowdeg oracle} implies the following:
\begin{prop}
\label{prop:bound time oracle lowdeg}Both preprocessing and query time of $\match^{\lowdeg}(\cdot)$ is at most $\Otil((t_{A}n^{2\epsilon}+n+t_{A}/\sizeOut)n^{2\epsilon}/\sizeOut^{2})=\Otil((t_{A}+n)n^{4\epsilon}/\sizeOut^{3})$ with high probability.
\end{prop}

\begin{lem}
\label{lem:prep time}\Cref{alg:prep} takes $\Otil((t_{A}+n)(n^{1-\epsilon}+n^{4\epsilon})/\poly(\sizeIn))$ total running time.
\end{lem}

\begin{proof}
We will analyze the total running time for each iteration of the repeat-loop in \Cref{alg:prep}. Since there are $T=O(1/\sizeIn^{2})$ iterations and we assume $\sizeIn\ge1/\poly\log n$, the running time is the same up to polylogarithmic factor. Now, fix one iteration of the repeat-loop. 

The total time to compute $M^{i}$, for all $i\le k$, is $\Theta(kp)=\Otil(n^{2-\epsilon})$. For each for-loop iteration, to compute $\tilde{\mu}_{1}$, we make $\Theta(r_{1})$ queries to $\mem_{A}$ taking $\Theta(r_{1})\cdot t_{A}=\Otil(t_{A}/\sizeIn^{10})$ time. To compute $\tilde{\mu}_{2}$, we make $r_{2}$ queries to $\match^{\lowdeg}(\cdot)$. By \Cref{prop:bound time oracle lowdeg}, this takes time $r_{2}\cdot\Otil((t_{A}+n)n^{4\epsilon}/\sizeOut^{3})=\Otil((t_{A}+n)n^{4\epsilon}/\sizeIn^{25})$ by \Cref{lem:lowdeg oracle}.

Next, we analyze the time to compute $A'_{\sp}$. Since $|A'|\ge\sizeIn n$ w.h.p. by \Cref{lem:large matching remain}, we can sample a random vertex in $A'$ by sampling at most $O(\log n/\sizeIn)$ times in $V'$ w.h.p. For each sample, we need to make a query to $\mem_{A}$, so we can compute $A'_{\sp}$ in time $O(r_{3})\cdot O(t_{A}\log n/\sizeIn)=\Otil(t_{A}n^{1-\epsilon}/\poly(\sizeIn))$ because $r_{3}=1000\sizeIn\frac{n}{k}\log n$ and $k=n^{\epsilon}$. Once $A'_{\sp}$ is computed, we can compute $C$ in $|E(\Gbar)|=\Theta(kp)=\Otil(n^{2-\epsilon})$. To conclude, the total running time in each iteration of the repeat-loop at most 
\[
\Otil(n^{2-\epsilon}+(t_{A}+n)n^{4\epsilon}+t_{A}n^{1-\epsilon})/\poly(\sizeIn)=\Otil((t_{A}+n)(n^{1-\epsilon}+n^{4\epsilon})/\poly(\sizeIn)).
\]
\end{proof}
The main lemma on preprocessing, \Cref{lem:prep}, is implied by combining \Cref{cor:prep correct}, \Cref{cor:conclude terminate} and \Cref{lem:prep time}

\subsection{Query Algorithm}

\label{sec:induced query}

We define our matching oracle $\match$ depending on the cases from \Cref{lem:prep}.

Suppose \Cref{lem:prep} returns $M'$ that satisfies Case~\ref{enu:explicit}. Let $M_{1}=M'[A]$. By \Cref{lem:prep}, $|M_{1}|\ge2\sizeOut n$. The algorithm for outputting $\match(v)$ with respect to $M_{1}$ is described in \Cref{alg:query1}. The correctness is straightforward and the worst-case query time is clearly $2t_{A}+O(1)$.

\begin{algorithm}
\begin{enumerate}
\item If $v\in V(M')$, let $v'$ be such that $(v,v')\in M'$. Else, return $\bot$.  
\item If $\mem_{A}(v),\mem_{A}(v')=1$, return $(v,v')$. Else, return $\bot$. 
\end{enumerate}
\caption{Compute $\protect\match(v)$ with respect to $M_{1}$.\label{alg:query1}}
\end{algorithm}

Next, suppose \Cref{lem:prep} returns $M'$ that satisfies Case~\ref{enu:implicit}. Let $M_{2}$ be a $(2,\sizeOut n)$-approximate matching in $G[A\cap V'\setminus V(M')]$. By \Cref{lem:prep}, $|M_{2}|\ge\mu(G[A\cap V'\setminus V(M')])/2-\sizeOut n\ge\sizeOut n$.\footnote{In fact, if we define $M_{2}$ as $\Mhat^{i}$ from Line~\ref{enu:define lowdeg matching} in \Cref{alg:prep}, we would even have that $|M_{2}|\ge4\sizeIn n$ w.h.p. But we did use this bound just to avoid white-boxing the preprocessing algorithm and make the presentation of the query algorithm more modular.} The algorithm for outputting $\match(v)$ with respect to $M_{2}$ is described in \Cref{alg:query2}. The correctness is straightforward. Let us analyze the query time. Step 1 takes $t_{A}+O(1)$ time. Step 2 takes $\Otil((t_{A}+n)n^{4\epsilon}/\sizeOut^{3})$ following the same proof as in \Cref{prop:bound time oracle lowdeg} (the maximum degree of $G[V'\setminus V(M')]$ is at most $n^{2\epsilon}$ w.h.p. by \Cref{lem:sparsification}).

\begin{algorithm}
Let $\match_{\lowdeg}$ be the $(2,\sizeOut n)$-approximate matching oracle from \Cref{lem:lowdeg oracle} when given graph $G[V'\setminus V(M^{i})]$ with vertex set $A\cap V'$ as input.
\begin{enumerate}
\item Check if $v\in A\cap V'\setminus V(M')$. If not, return $\bot$.
\item Using the oracle $\match_{\lowdeg}$, if $v\in V(M_{2})$, return $(v,v')\in M_{2}$. Else, return $\bot$. 
\end{enumerate}
\caption{Compute $\protect\match(v)$ with respect to $M_{2}$.\label{alg:query2}}
\end{algorithm}

In both cases, the matching oracle $\match$ respects a matching of size at least $\sizeOut n$ and has worst-case query time at most $\tilde{O}((t_{A}+n)n^{4\epsilon}/\poly(\sizeIn))$ w.h.p.

%% file: 5-boosting.tex
\newcommand{\mat}{\texttt{matrix}}

\renewcommand{\mem}{\texttt{mem}}

\newcommand{\alive}{\texttt{alive}}
\newcommand{\template}{\texttt{Augment-Template}}

\renewcommand{\P}{\mathcal{P}}

\newcommand{\M}{\mathcal{M}}

\section{Boosting the Approximation Guarantee of a Matching Oracle}
\label{sec:mcgregor}
\label{sec:boost}

Recall the notations from Section~\ref{sec:prelim}. Throughout this section, we use the following parameters.

\begin{definition}
\label{def:parameters}
$k \geq 0$ is an integral constant, $\gamma \in (0, 1)$ is a constant, $T = \Theta_{k, \gamma}(1)$ is a sufficiently large integral constant that depends only on  $k$  and  $\gamma$ (see Lemma~\ref{lm:template}), and $\epsilon_{\inp} > 0$ is a sufficiently small constant such that $9^T \cdot \epsilon_{\inp} < 1/5$.
\end{definition}

We present an algorithm $\algo(G, M^{\inp}, k, \gamma, \epsilon_{\inp})$, which takes as input: a graph $G = (V, E)$ with $n$ nodes, the parameters $k, \gamma, \epsilon_{\inp}$ as in Definition~\ref{def:parameters},  and an oracle $\match_{M^{\inp}}(.)$ for a matching $M^{\inp}$ in $G$ that has   $\tilde{O}_{k, \gamma}(n^{1+\epsilon_{\inp}})$ query time.
The algorithm either returns an oracle $\match_{M^{\out}}(.)$ for a matching $M^{\out}$ in $G$ that is obtained by applying a sufficiently large number of length $(2k+1)$-augmenting paths to $M^{\inp}$, or it returns {\sc Failure}. We now state our main result in this section.

\begin{thm}
\label{thm:boost} 
Set $\epsilon_{\out} := 9^T \cdot \epsilon_{\inp}$ (see Definition~\ref{def:parameters}). Given adjacency-matrix query access to the input graph $G = (V, E)$, the algorithm $\algo(G, M^{\inp}, k, \gamma, \epsilon_{\inp})$ runs in $\tilde{O}_{k, \gamma}\left(n^{2-\epsilon_{\inp}}\right)$ time. Further,  either it returns an oracle $\match_{M^{\out}}(.)$ with query time $\tilde{O}_{k,\gamma}(n^{1+\epsilon_{\out}})$, for some matching $M^{\out}$ in $G$ of size $|M^{\out}| \geq |M^{\inp}| + \Theta_{k, \gamma}(1) \cdot n$ (we say that it ``succeeds'' in this case), 
 or it returns {\sc Failure}.  Finally, if the matching $M^{\inp}$ admits a collection of $\gamma \cdot n$ many node-disjoint length $(2k+1)$-augmenting paths in $G$, then the algorithm succeeds whp.
\end{thm}

In Section~\ref{sec:algo:description}, we present a {\em template algorithm} for the task stated in Theorem~\ref{thm:boost}. This  is inspired by  an algorithm of McGregor~\cite{mcgregor2005finding} for computing a $(1+\epsilon)$-approximate  matching in the semi-streaming model.
While describing the template algorithm, we assume that we are given the matching $M^{\inp}$ explicitly as part of the input, and that we need to either construct the matching $M^{\out}$ or return {\sc Failure}. Note, however, that in the sublinear setting, we cannot assume this.

Subsequently, in Section~\ref{sec:sublinear:implementation}, we show how to implement the template algorithm in the sublinear setting under adjacency-matrix queries, which leads to the proof of Theorem~\ref{thm:boost}.
  
 \paragraph{Remark on Oracles:} Throughout this section, we will treat the oracle $\match_M(.)$ as a data structure in the sublinear model, which returns the appropriate answer upon receiving a query. In contrast, we will treat the oracle $\mate_M(.)$ as simply an abstract function, so that $\mate_M(v)$ simply denotes the mate of $v$ (if it exists) under  $M$ (see Section~\ref{sec:prelim}). Note that we can return the value of $\mate_M(v)$ by making a single query to  $\match_M(v)$, without any additional overhead in time.

  \subsection{A Template Algorithm}
  \label{sec:algo:description}

We denote the template algorithm simply by $\template(G, M^{\inp}, k, \gamma)$, as we do not need the parameter $\epsilon_{\inp}$ to describe it. The parameter $\epsilon_{\inp}$ will become relevant only in Section~\ref{sec:sublinear:implementation}, when we consider implementing this algorithm in the sublinear setting.

As part of the input to the template algorithm, the $n$-node graph $G = (V, E)$ and the matching $M^{\inp}$ are specified explicitly. The algorithm either returns an explicit matching $M^{\out}$  in $G$ of size $|M^{\out}| \geq |M^{\inp}| + \Theta_{k, \gamma}(1) \cdot n$ (we say that it ``succeeds'' in this case),
or it returns {\sc Failure}. If $M^{\inp}$ admits a collection of $\gamma \cdot n$ many node-disjoint length $(2k+1)$-augmenting paths in $G$, then the template algorithm succeeds whp. This mimics  Theorem~\ref{thm:boost}. Furthermore,   the template algorithm has access to a subroutine $\largematch(S, \delta)$, which takes as input  a subset of nodes $S \subseteq V$ and a small constant $\delta \in (0, 1)$, and either returns $\perp$ or returns a matching $M$ in  $G[S]$ such that $|M| \geq \frac{1}{10^8} \cdot \delta^5 \cdot n$. In addition, if $\mu(G) \geq \delta \cdot n$, then it is guaranteed that $\largematch(G, \delta)$ does {\em not}
 return $\perp$. This mimics   Theorem~\ref{thm:matching oracle}, with $\delta_{\inp} = \delta$.

\subsubsection{Algorithm Description}
\label{sec:template:describe}

\noindent {\bf Random partitioning:}
  We start by partitioning the node-set $V$ into $2k+2$ subsets $L_0, \ldots, L_{2k+1}$, as follows. For each $v \in V$, we place the node $v$ into one of the subsets $L_0, \ldots, L_{2k+1}$ chosen uniformly and independently at random. We will refer to the subset $L_{i}$ as {\em layer} $i$ of this partition. If  $v \in L_i$, then we will write $\ell(v) = i$ and simply say that the node $v$ {\em belongs to layer $i$}.

\medskip
Let $p$ be an augmenting path of length $(2k+1)$ in $G$ w.r.t.~$M^{\inp}$. Assign an arbitrary direction to this path, so that we can write $p = (v_0, v_1, \ldots, v_{2k+1})$ w.l.o.g. Specifically, we have $(v_{2i}, v_{2i+1}) \in E \setminus M^{\inp}$ for all $i \in [0, k]$, and $(v_{2i-1}, v_{2i}) \in M^{\inp}$ for all $i \in [1, k]$. We say that the path $p$ {\em survives} the random partitioning iff $v_i \in L_i$ for all $i \in [0, 2k+1]$.

\begin{lem}
\label{lm:survive}
Consider any collection $\P$ of node-disjoint length $(2k+1)$-augmenting paths in $G$ w.r.t.~$M^{\inp}$. Let $\P^* \subseteq \P$ denote the subset of paths in $\P$ that survive the random partitioning. If $|\P| \geq \gamma \cdot n$, then $|\P^*| \geq \Theta_{k, \gamma}(1) \cdot n$ whp.
\end{lem}

\begin{proof}
Each path $p \in \P$ survives the random partitioning with probability $(2k+2)^{-(2k+2)}$. Since $|\P| \geq \gamma \cdot n$, by linearity of expectation, we get: $\E[| \P^* |] \geq \left( (2k+2)^{-(2k+2)} \gamma\right) \cdot n = \Theta_{k, \gamma}(1) \cdot n$. Finally, we note that whether a given path $p \in \P$ survives the random partitioning or not is independent of the fate of the other paths in $\P$. The lemma now follows from a Chernoff bound.
\end{proof}

Motivated by Lemma~\ref{lm:survive}, the template algorithm will only attempt to augment $M^{\inp}$ along those augmenting paths  that survive the random partitioning. This leads us to introduce the notion of {\em layered subgraphs} of $G$, as described below. Intuitively,  although the template algorithm does not know the set $\P^*$ in advance, it can be certain that the sequence of edges in any length $(2k+1)$-augmenting path in $\P^*$ appears in successive layered subgraphs (see Observation~\ref{obs:layering:2}).

\paragraph{Layered subgraphs of $G$:} First, we define a set  $V_H \subseteq V$.  A node $v \in V$ belongs to $V_H$ iff either
\begin{enumerate}
    \item $\ell(v) \in \{0, 2k+1\}$ and $\mate_{M^{\inp}}(v) = \perp$, 
    or
    \item $\ell(v) = 2j - 1$ for some $j \in [1, k]$ and $\ell\left(\mate_{M^{\inp}}(v) \right) = 2j$, 
    or
    \item $\ell(v) = 2j$ for some $j \in [1, k]$ and $\ell\left( \mate_{M^{\inp}}(v)\right) = 2j-1$. 
\end{enumerate}
Given the nodes in $V_H$, the edge-set $E_H \subseteq E$ is defined as follows. An edge $(u, v) \in E$ belongs to $E_H$ iff 
$u, v \in V_H$, $\left| \ell(u) -\ell(v)\right| = 1$, and 
either 
\begin{enumerate}
    \item $\min(\ell(u), \ell(v))$ is even and $(u, v) \notin M^{\inp}$, or
    \item $\min(\ell(u), \ell(v))$ is odd and $(u, v) \in M^{\inp}$    
\end{enumerate}
We next define the subgraph $H := (V_H, E_H)$.  Finally, for each $i \in [0, 2k]$, let  $G_i := (V, E_i)$ be a bipartite subgraph of $G$, where $E_i := \{ (u, v) \in E_H : \ell(u) = i, \ell(v) = i+1\}$ is the set of edges in $E_H$  between layer $i$ and layer $i+1$. Note that we have defined the subgraphs $\{G_i\}$ over the entire node-set $V$, although every edge in these subgraphs has both its endpoints in $V_H$. This is done to simplify notations, as will become evident when we describe how to implement our algorithm in the sublinear setting. For each $i \in [0, 2k+1]$, we refer to the nodes in $V_{i} := L_i \cap V_H$ as being {\em relevant} for the concerned layer.

\medskip
We now state some key observations, which immediately follow from the description above.

\begin{obs}
\label{obs:layering:0}
For all $i \in [1, 2k]$, we have $V_i \subseteq V(M^{\inp})$.
\end{obs}

 \begin{obs}
\label{obs:layering:1}
For all $i \in [0, k]$, we have $E_{2i} = E\left( G[V_{2i} \cup V_{2i+1}]\right)$. Furthermore, for all $i \in [1, k]$, we have $E_{2i-1} = M^{\inp} \cap \left( V_{2i-1} \times V_{2i}\right)$. Thus, if $i$ is even, then $G_i$  consists of all the edges from $G$ that connect two relevant nodes across the concerned layers. In constrast, if $i$ is odd, then $G_i$ consists of  the edges from $M^{\inp}$ that connect two relevant nodes across the concerned layers.
 \end{obs}

\begin{obs}
\label{obs:layering:2} Consider any augmenting path $p = (v_0, v_1, \ldots, v_{2k+1})$ w.r.t.~$M^{\inp}$ in $G$ that survives the random partitioning. Then we have $(v_{i}, v_{i+1}) \in E_i$ for all $i \in [0, 2k]$.
\end{obs}

 \paragraph{Nested matchings:} Fix any $j \in [0, k]$, and for each $i \in [0, j]$ consider a matching $M_{2i} \subseteq E_{2i}$ in $G_{2i}$. We say that the sequence of matchings $M_0, M_2, \ldots M_{2j}$ is {\em nested} iff for all $i \in [1, j]$ and all $v \in V(M_{2i}) \cap V_{2i}$, we have $\mate_{M^{\inp}}(v) \in V(M_{2i-2})$.

 \begin{obs}
 \label{new:obs:nested}
 Consider any sequence of nested matchings $M_0, M_2, \ldots, M_{2k}$.  Then there exists a collection of node-disjoint length $(2k+1)$-augmenting paths of size $\left| M_{2k} \right|$ w.r.t.~$M^{\inp}$ in $G$.
 \end{obs}

\begin{proof}
Consider any node $v \in V(M_{2k}) \cap V_{2k+1}$. Consider the  path $p(v) = (v_0, v_1, \ldots, v_{2k+1})$ in $G$, which is constructed according to the following procedure.
\begin{itemize}
\item $v_{2k+1} \leftarrow v$, and $i \leftarrow 2k$. (Note that $v_{2k+1}$ is at layer $2k+1$ and is matched under $M_{2k}$.)
\item {\sc While} $i \geq 0$:
\begin{itemize}
\item If $i$ is even, then  $v_i \leftarrow \mate_{M_i}(v_{i+1})$.
\item If $i$ is odd, then  $v_i \leftarrow \mate_{M^{\inp}}(v_{i+1})$.
\item $i \leftarrow i-1$.
\end{itemize}
\end{itemize}
Since the sequence  $M_0, M_2, \ldots, M_{2k}$ is nested, applying Observation~\ref{obs:layering:0} and Observation~\ref{obs:layering:1}, we can show (by an induction on the number of iterations of the {\sc While} loop) that $p(v)$ is a length $(2k+1)$-augmenting path in $G$ w.r.t.~$M^{\inp}$. Furthermore, it is easy to see that the paths $\{ p(v) \}_{v \in V(M_{2k}) \cap V_{2k+1}}$ constructed in this manner are mutually node-disjoint. This implies the observation.
\end{proof}

\paragraph{Important parameters:} 
We fix a constant $\psi \in \left(0, 1\right)$, which  depends on $k$ and $\gamma$, i.e., $\psi = \Theta_{k,\gamma}(1)$,  and is chosen to be sufficiently small (see Corollary~\ref{cor:bound:backtrack}). Next, for each $i \in [0, k]$, we define:
\begin{equation}
\label{eq:define:delta}
\psi_{i} := \frac{1}{10^8} \cdot \psi^{5^{4i+3}} \text{ and } \delta_i := \psi^{5^{4i+1}}. 
\end{equation}

Consider any matching $M'$ between the nodes at layers $2i$ and $2i+1$, where $i \in [0, k]$. Intuitively, the parameters $\psi_i$ and $\delta_i$ will determine how large $M'$ needs to be so as to make us ``happy''.
Note that the values of $\delta_{i}$ and $\psi_i$ decrease in a doubly exponential manner with $i$. This fact will be crucially used during the analysis in Section~\ref{sec:template:analyze}.

 \paragraph{A relatively informal summary of the algorithm:}
Motivated by Observation~\ref{new:obs:nested}, the template algorithm attempts to find a sequence of nested matchings ending at layer $2k$. Specifically, the algorithm runs in {\em iterations}. At the start of a given iteration, we maintain a sequence of nested matchings $M_0, M_2, \ldots, M_{2i}$ up to some layer $2i$, such that $|M_{2j}| \geq \psi_{j} \cdot n$ for all $j \in [0, i]$. If $i = k$, then by Observation~\ref{new:obs:nested} we can already identify a collection of $\psi_k \cdot n = \Theta_{k, \gamma}(1) \cdot n$ many node-disjoint length $(2k+1)$-augmenting paths in $G$ w.r.t.~$M^{\inp}$, and so we just apply those augmenting paths to $M^{\inp}$ and return the resulting matching $M^{\out}$. Henceforth, assume that $i < k$. We classify each node in $V_H$ as either {\em alive} or {\em dead} (at the start of the first iteration every node was alive). We also enforce the invariant that  all the nodes currently matched in $M_0 \cup M_2 \cup \cdots \cup M_{2i}$ are alive. During the current iteration, we attempt to find a large matching $M'$ between the alive nodes in $G_{2i+2}$, while ensuring that the sequence $M_0, M_2, \ldots, M_{2i}, M'$ remains nested. Specifically, we make a call to the subroutine $\largematch(S,\delta_{i+1})$,  for an appropriate $S \subseteq V_{2i+2} \cup V_{2i+3}$. Depending on the outcome of this call, we now fork into one of the following three cases.
\begin{description}
\item[Case (a):] The call to $\largematch(S, \delta_{i+1})$ returns a matching $M'$. Thus, we are guaranteed that $|M'| \geq \frac{1}{10^8} \cdot (\delta_{i+1})^5 \cdot  n \geq \psi_{i+1} \cdot n$. We set $M_{2i+2} := M'$, $i := i+1$, and proceed to the next iteration.
\item[Case (b):] The call to $\largematch(S, \delta_{i+1})$ returns $\perp$, and $i= -1$ (i.e., the sequence of matchings $M_0, M_2, \ldots, M_{2i}$ was empty). Here, we terminate the template algorithm and return {\sc Failure}.
\item[Case (c):] The call to $\largematch(S, \delta_{i+1})$ returns $\perp$, and $i \geq 0$. Here, we change the status of all the nodes in $V(M_{2i}) \cap V_{2i+1}$, along with their matched neighbors under $M^{\inp}$ (who are at layer $2i+2$), from alive to dead. We then delete the matching $M_{2i}$, set $i := i - 1$, and proceed to the next iteration.
\end{description}

We will need some more notations while working with this algorithm in Section~\ref{sec:sublinear:implementation}. Accordingly, below we present a more detailed and technical description of the template algorithm, along with these additional notations. While going through the rest of this section, the reader will find it helpful to refer back to the informal description above, whenever necessary.

\paragraph{Iterations:}  In each iteration $t \geq 1$, we will compute a matching $M^{(t)}$ in the subgraph $G_{\sigma(t)}$, where $\sigma(t) \in \{0, 2, 4, \ldots, 2k\}$. The mapping $\sigma : T \rightarrow \{0, 2, \ldots, 2k\}$ will be constructed in an online manner, i.e., we will assign the value $\sigma(t)$ only during the $t^{th}$ iteration.  We now describe the state of the algorithm at the end of any given iteration.

At the end of an iteration $t$, a subset of past iterations $\Lambda^{(t)} \subseteq [t]$ are designated as being {\em active} w.r.t.~$t$.  If $\Lambda^{(t)} \neq \emptyset$, then we  write $\Lambda^{(t)} := \left\{ \lambda^{(t)}_0, \lambda^{(t)}_1, \ldots, \lambda^{(t)}_{\stack(t)} \right\}$, where $\stack(t) := \left| \Lambda^{(t)} \right| - 1$ and
 $\lambda^{(t)}_0 < \lambda^{(t)}_1 < \cdots < \lambda^{(t)}_{\stack(t)}$. The sequence of matchings $M^{\left(\lambda^{(t)}_0\right)}, M^{\left(\lambda^{(t)}_1\right)}, \ldots, M^{\left(\lambda^{(t)}_{\stack(t)}\right)}$ corresponds to the sequence $M_0, M_2, \ldots, M_{2i}$ in the discussion immediately after Observation~\ref{new:obs:nested}. Thus, the algorithm satisfies the following invariants. 
 \begin{invariant}
 \label{new:inv:1} We have $\stack(t) \leq k$, and $\sigma\left( \lambda^{(t)}_j \right) = 2 j$ for all $j \in \left[0, \stack(t) \right]$. 
 \end{invariant}

 \begin{invariant}
 \label{new:inv:nested}
 The sequence of matchings $M^{\left(\lambda^{(t)}_0\right)}, M^{\left(\lambda^{(t)}_1\right)}, \ldots, M^{\left(\lambda^{(t)}_{\stack(t)}\right)}$ is nested. 
 \end{invariant}
 
 \begin{invariant}
 \label{new:inv:size}
 $\left| M^{\left(\lambda^{(t)}_j\right)} \right| \geq \psi_{j} \cdot n$ for all $j \in \left[0,  \stack(t)\right]$. 
 \end{invariant}
 
 For each layer $i \in [0, 2k+1]$, the set of relevant nodes $V_i$ is partitioned into two subsets: $A_i$ and $D_i$. We refer to the nodes in $A_i$ as {\em alive}, and the nodes in $D_i$ as {\em dead}. We let $A := \bigcup_{i = 0}^{2k+1} A_i$ and $D := \bigcup_{i=0}^{2k+1} D_i$ respectively denote the set of all alive and dead nodes, across all the layers. The next invariant states that every matched node in an active iteration is alive.
 
 \begin{invariant}
 \label{new:inv:alive}
 $A \supseteq  V \left( M^{\left(\lambda^{(t)}_j\right)}\right)$ for all $j \in \left[0,  \stack(t) \right]$. 
 \end{invariant}
 
 At the start of the first iteration (when $t = 1$), every relevant node is alive (i.e., $A_i = V_i$ and $D_i = \emptyset$ for all $i \in [0, 2k+1]$). Subsequently, over time the status of a relevant node  can only change from being alive to being dead, but {\em not} the other way round. Thus, with time, the set $D$ keeps growing, where the set $A$ keeps shrinking. We now explain how to implement a given iteration $t$.

 \paragraph{Implementing iteration $t$:} Let $i = \stack(t-1)$. 
 If $\Lambda^{(t-1)} = \emptyset$, then we set  $i = -1$. If $i = k$, then there will be no more iterations, i.e., the algorithm will last for only $t-1$ iterations. In this scenario, we know that the sequence of matchings $M^{\left(\lambda^{(t-1)}_0\right)}, M^{\left( \lambda^{(t-1)}_1\right)},\cdots , M^{\left( \lambda^{(t-1)}_k\right)}$ is nested.  Based on this sequence, we identify a collection of  $| M^{\left( \lambda^{(t-1)}_k \right)}|$ many node-disjoint augmenting paths w.r.t.~$M^{\inp}$ in $G$,  augment $M^{\inp}$ along those paths (see Observation~\ref{new:obs:nested}), and return the resulting matching $M^{\texttt{out}}$. Accordingly, from now on we assume that $i \leq k-1$.

 During iteration $t$, we will attempt to find a large matching $M'$ in $G_{2i+2}$ between two sets of nodes: $A_{2i+3}$ and $C_{2i+2}$. Recall that $A_{2i+3}$ denotes the alive nodes at layer $2i+3$. We refer to $C_{2i+2}$ as the set of {\em candidate nodes} for iteration $t$. Intuitively, we pick as many nodes from $V_{2i+2}$ into the set $C_{2i+2}$ as possible, subject to  two constraints: (i) if we append $M'$ at the end of the sequence of matchings from the currently active iterations, then the resulting sequence will continue to remain nested, and (ii) the nodes in $C_{2i+2}$ are currently alive. This leads us to the following definition.
 \begin{equation*}
C_{2i+2} := \begin{cases}  
\left\{ v \in A_{2i+2} : \mate_{M^{\inp}}(v)  \in V\left( M^{\left(\lambda^{(t-1)}_{i}\right)}\right) \right\} & \text{ if } i \geq 0; \\
 A_{2i+2} & \text{ else if } i = -1.
\end{cases}
 \end{equation*}

 We now call the subroutine $\largematch(C_{2i+2} \cup A_{2i+3}, \delta_{i+1})$, in an attempt to  obtain a large matching in $G\left[C_{2i+2} \cup A_{2i+3}\right]= G_{2i+2}\left[C_{2i+2} \cup A_{2i+3}\right]$. The last equality holds because of Observation~\ref{obs:layering:1}, and since $C_{2i+2} \subseteq V_{2i+2}$ and $A_{2i+3} \subseteq V_{2i+3}$. We set $\sigma(t) := 2i+2$. Now, we fork into one of the following three cases.
 
\begin{description}
 \item[Case (a):] The call to $\largematch(C_{2i+2} \cup A_{2i+3}, \delta_{i+1})$ returns a matching $M'$. Thus, we are guaranteed that $|M'| \geq \frac{1}{10^8} \cdot (\delta_{i+1})^5 \cdot n \geq \psi_{i+1} \cdot n$. We set $M^{(t)} := M'$, $\Lambda^{(t)} := \Lambda^{(t-1)} \cup \{ t \}$ and $\stack(t) := \stack(t-1) +1$.   This implies that  $\lambda^{(t)}_j := \lambda^{(t-1)}_j$ for all $j \in \left[0, \stack(t-1)\right]$, and $\lambda^{(t)}_{\stack(t)} := t$. Henceforth, we  refer to this iteration $t$ as a {\em forwarding iteration} at layer $(2i+2)$. We now move on to  the next iteration $(t+1)$.
\item[Case (b):] The call to $\largematch(C_{2i+2} \cup A_{2i+3}, \delta_{i+1})$ returns  $\perp$, and $i = -1$.
Here, the algorithm terminates and returns {\sc Failure}. 
 \item[Case (c):] The call to $\largematch(C_{2i+2} \cup A_{2i+3}, \delta_{i+1})$ returns  $\perp$, and $i \geq 0$. Here, we set $M^{(t)} := \emptyset$.  We also change the status of all the nodes in $C_{2i+2}$,   along with their matched neighbors under $M^{\inp}$ (who are at layer $2i+1$), from alive to dead, and respectively move these nodes from $A_{2i+1}$ to $D_{2i+1}$ and from $A_{2i+2}$ to $D_{2i+2}$. Next, we set $\Lambda^{(t)} := \Lambda^{(t-1)} \setminus \{ \lambda^{(t-1)}_{i} \}$ and $\stack(t) := \stack(t-1) - 1$. This implies that $\lambda^{(t)}_j := \lambda^{(t-1)}_j$ for all $j \in \left[0, \stack(t)\right]$. Henceforth, we  refer to this iteration $t$ as a {\em backtracking iteration} for layer $2i$. We now move on to the next iteration $(t+1)$.
 \end{description}

\paragraph{Remark:} From the above description of the template algorithm, it immediately follows that Invariants~\ref{new:inv:1},~\ref{new:inv:nested},~\ref{new:inv:size} and~\ref{new:inv:alive} continue to hold at the end of each iteration $t$.

\subsubsection{Analysis}
\label{sec:template:analyze}

In this section, we analyze the template algorithm, and prove the following lemma.

\begin{lem}
\label{lm:template}
The algorithm $\template(G, M^{\inp}, k, \gamma)$ runs for at most $T = \Theta_{k, \gamma}(1)$ iterations. It either returns a matching $M^{\out}$ in $G$ of size $|M^{\out}| \geq |M^{\inp}| + \Theta_{k, \gamma}(1) \cdot n$ (we say that the algorithm ``succeeds'' in this case), or it returns {\sc Failure}.  Furthermore, if $M^{\inp}$ admits a collection of $\gamma \cdot n$ many node-disjoint length $(2k+1)$-augmenting paths in $G$, then the algorithm succeeds whp.
\end{lem}

We start by focusing on bounding the number of iterations (see Corollary~\ref{cor:backtracking}).

\begin{claim}
\label{cl:backtracking}
 There can be at most $1/(\psi_{i})$  backtracking iterations for layer $2i$, where $i \in [0, k-1]$.
\end{claim}

\begin{proof}
Consider any backtracking iteration $t$ for layer $2i$. Then we have $\sigma(t-1) = i$, and Invariant~\ref{new:inv:size} implies that $$\left| V\left(M^{\left( \lambda^{(t-1)}_{i} \right)} \right) \cap V_{2i+1}\right| = \left| M^{\left( \lambda_i^{(t-1)}\right)} \right| \geq \psi_i \cdot n.$$ Thus, during iteration $t$, at least $\psi_i \cdot n$ nodes at layer $(2i+1)$ change their status from alive to dead. Since there are at most $n$ nodes at layer $(2i+1)$, such an event can occur at most $1/(\psi_{i})$ times.
\end{proof}

\begin{cor}
\label{cor:backtracking}
The algorithm $\template(G, M^{\inp}, k, \gamma)$ has at most $\Theta_{k, \gamma}(1)$ iterations. 
\end{cor}

\begin{proof}
Let $T_{f}, T_{b}$ and $T_{0}$ respectively denote the total number of forwarding iterations across all layers, the total number of backtracking iterations across all layers, and the total number of iterations across all layers that are neither forwarding nor backtracking.
We have $T_0 = 1$ if the template algorithm returns {\sc Failure}, and $T_0 = 0$ otherwise. 

We now observe that: there cannot exist a sequence of more than $(k+1)$ consecutive forwarding iterations, for otherwise, the $(k+2)^{th}$ forwarding iteration in this sequence would have to take place at a layer $\geq (2k+2)$, which does not exist. Hence, we have: $T_f \leq (k+1) \cdot (T_b+T_0) + (k+1)$, and the total number of iterations  is bounded by:
$$T = T_f + T_b + T_0 \leq (k+1) \cdot (T_b + T_0) + (k+1) + T_b + T_0  = \Theta(k) \cdot T_b \leq \Theta(k) \cdot  \sum_{i=0}^{k-1} \frac{1}{\psi_{i}}  = \Theta_{k, \gamma}(1).$$
The second inequality follows from Claim~\ref{cl:backtracking}, and the last equality follows from~(\ref{eq:define:delta}).
\end{proof}

We now move on to showing that if $M^{\inp}$ admits a collection $\gamma \cdot n$ many node-disjoint length $(2k+1)$-augmenting paths in $G$, then  the template algorithm succeeds whp.  Towards this end, let $\P$ denote a maximum-sized collection of node-disjoint length $(2k+1)$-augmenting paths in $G$ w.r.t.~$M^{\inp}$. Let $\P^* \subseteq \P$ be the subset of paths in $\P$ that survive the random partitioning. If $|\P| \geq \gamma \cdot n$, then Lemma~\ref{lm:survive} guarantees that whp:
 \begin{equation}
\label{eq:whp}
|\P^*| \geq \Theta_{k, \gamma}(1) \cdot n.
\end{equation}

At any point in time during the execution of the algorithm $\template(G, M^{\inp}, k, \gamma)$, we say that a path $p \in \P^*$ is {\em alive} if {\em all} the nodes on $p$ are alive, and we say that the path $p$ is {\em dead} otherwise. Let $\P^*_A \subseteq \P^*$ and $\P^*_D = \P^* 
\setminus \P^*_A$ respectively denote the set of alive and dead paths at any point in time. Just before the start of iteration $1$, we have $\P^*_A = \P^*$ and $\P^*_D = \emptyset$. Subsequently, a path $p \in \P^*$ can change its status from alive to dead only during a backtracking iteration (note that this change occurs in only one direction, i.e., a dead path will never become alive). The next claim upper bounds the number of such changes.

\begin{claim}
\label{cl:bound:backtrack}
During a backtracking iteration for layer $2i$, where $i \in[0, k-1]$, at most $\delta_{i+1} \cdot n$ many paths in $\P^*$ moves from $\P^*_A$ to $\P^*_D$.
\end{claim}

\begin{proof}
    Let $t \geq 1$ denote a  backtracking iteration for layer $2i$. During iteration $t$, the algorithm calls $\largematch(C_{2i+2} \cup A_{2i+3}, \delta_{i+1})$, which returns $\perp$. Consider the subgraph $G' = G[C_{2i+2} \cup A_{2i+3}]$.  We have: $\mu(G') < \delta_{i+1} \cdot n$, 
    for otherwise the call to $\largematch(., .)$ would not have returned $\bot$.

  Just before iteration $t$,  let $\P' \subseteq \P^*_A$ denote the subset of paths in $\P^*_A$ that pass through some node in $C_{2i+2}$. Only the paths in $\P'$ move from $\P^*_A$ to $\P^*_D$ at the end of iteration $t$. We can, however, form a matching in $G'$ which contains one edge from each path in $\P'$. Hence, we have $|\P'| \leq \mu(G') < \delta_{i+1} \cdot n$. This concludes the proof of the claim.
\end{proof}

\begin{cor}
\label{cor:bound:backtrack}
Let $\psi = \Theta_{k, \gamma}(1)$ be a sufficiently small constant depending on $k$ and $\gamma$, and suppose that~(\ref{eq:whp}) holds. Then throughout the entire duration of the algorithm, we have:
$$\left| \P^*_{A} \right| \geq \left| \P^* \right| -  \sum_{i=0}^{k-1}\frac{\delta_{i+1}}{\psi_{i}} \cdot n \geq \delta_0 \cdot n.$$
\end{cor}

\begin{proof}
From~(\ref{eq:define:delta}), Claim~\ref{cl:backtracking} and Claim~\ref{cl:bound:backtrack}, we infer that:
\begin{equation}
\label{eq:absorb}
\left| \P^*_A \right| \geq \left| \P^* \right| - \sum_{i=0}^{k-1}\frac{\delta_{i+1}}{\psi_{i}} \cdot n \geq \left| \P^* \right| - k \cdot (10^8 \psi) \cdot n.
\end{equation}
Now, since we can set $\psi$ to be any sufficiently small constant value depending on $k$ and $\gamma$, and since $\delta_0 \leq \psi$ according to~(\ref{eq:define:delta}), from~(\ref{eq:whp}) we get: $|\P^*| - k \cdot (10^8\psi) \cdot  n \geq \delta_0 \cdot n$. This concludes the proof. 
\end{proof}

\begin{cor}
\label{cor:terminate} If~(\ref{eq:whp}) holds, then the algorithm does {\em not} return {\sc Failure}.
\end{cor}

\begin{proof}
For contradiction, suppose that the algorithm returns {\sc Failure} at the end of an iteration $t$. 

Let $i = \sigma(t-1)$. Since the algorithm returns {\sc Failure} after iteration $t$, we must have $i = -1$. Furthermore, during iteration $t$, the call to $\largematch(C_0 \cup A_1, \delta_0)$ must have returned $\perp$. Let $G' = G[C_0 \cup A_1]$. It follows that:
\begin{equation}
\label{eq:contradict:500}
\mu(G') < \delta_0 \cdot n.
\end{equation}

Next, observe that $C_0 = A_0$. Hence, just before the start of iteration $t$, we could have formed a matching  in $G'$ by taking the first edge of each path in $\P^*_A$. Thus, from Corollary~\ref{cor:bound:backtrack}, we get: 
\begin{equation}
\label{eq:contradict:2}
\mu(G') \geq |\P^*_A| \geq   \delta_0 \cdot n.
\end{equation}
However, 
both~(\ref{eq:contradict:500}) and~(\ref{eq:contradict:2}) cannot simultaneously be true. This leads to a contradiction. 
\end{proof}

Note that if the template algorithm does {\em not} return {\sc Failure}, then it necessarily returns a matching $M^{\out}$ of size $|M^{\out}| \geq |M^{\inp}| + \psi_k \cdot n$ (this holds because of Invariant~\ref{new:inv:nested}, Invariant~\ref{new:inv:size} and Observation~\ref{new:obs:nested}). Finally, recall that $\psi_k = \Theta_{k, \gamma}(1)$ as per~(\ref{eq:define:delta}). Lemma~\ref{lm:template} now follows from Corollary~\ref{cor:backtracking}, Lemma~\ref{lm:survive} and Corollary~\ref{cor:terminate}.

\subsection{Implementation in Sublinear Models}
\label{sec:sublinear:implementation}

In this section, we show how to implement the template algorithm from Section~\ref{sec:algo:description}, when we are allowed access to the input graph $G$ only via adjacency-matrix queries. Throughout this section, we  use the following parameters (recall Definition~\ref{def:parameters}).
\begin{equation}
\label{eq:sublinear:reduction:1}
 \epsilon_0 := \epsilon_{\inp}, \text{ and } \epsilon_t := 9 \cdot \epsilon_{t-1} \text{ for all } t \in [1, T].
\end{equation}
In Section~\ref{sec:template:describe}, the template algorithm starts with iteration $t = 1$. Here,  we use the phrase ``iteration $t = 0$'' to refer to the scenario just before the start of the first iteration. Towards this end, for consistency of notations, we define $\epsilon_{-1} := 2$, $M^{(0)} := M^{\inp}$, $\sigma(0) := \perp$, $\stack(0) := -1$ and $\Lambda^{(0)} := \emptyset$. Further, we define an oracle $\alive_0(v)$ that is supposed to return {\sc Yes} if $v$ is alive at the end of iteration $0$ (i.e., just before the start of iteration $1$), and return {\sc No} otherwise. 

The rest of this section is organized as follows. Lemma~\ref{lm:sublinear:induction} shows how to implement each iteration of the template algorithm, under adjacency-matrix query access to the input graph $G$. Its proof appears at the end of this section.
Theorem~\ref{thm:boost} now follows from Lemma~\ref{lm:template} and Corollary~\ref{cor:sublinear:end}.

\begin{lem}
\label{lm:sublinear:induction}
Suppose that we can access the input graph $G$ only via adjacency-matrix queries, and we have an oracle $\match_{M^{\inp}}(.)$ with $\tilde{O}_{k, \gamma}(n^{1+\epsilon_{\inp}})$ query time. Then we can implement each iteration $t \geq 0$ of the algorithm $\template(G, M^{\inp}, k, \gamma)$, as described in Section~\ref{sec:algo:description}, in $\tilde{O}_{k, \gamma}(n^{2-\epsilon_{t-1}})$ time. Furthermore, if the concerned iteration $t$ does not result in the algorithm returning {\sc Failure}, then we can ensure that we have access to the following data structures at the end of iteration $t$.
\begin{itemize}
\item An oracle $\match_{M^{(t)}}(.)$ for the matching $M^{(t)}$, that has a query time of $\tilde{O}_{k, \gamma}(n^{1+\epsilon_t})$.
\item An oracle $\alive_t(.)$ that has a query time of $\tilde{O}_{k, \gamma}(n^{1+\epsilon_t})$. When queried with a node $v \in V$, this oracle returns {\sc Yes} if $v$ is alive at the end of iteration $t$, and returns {\sc No} otherwise.
\item The values $\sigma(t)$ and $\stack(t)$, and the contents of the set $\Lambda^{(t)}$.
\end{itemize}
\end{lem}

\begin{cor}
\label{cor:sublinear:end}
Let $\epsilon_{\out} := 9^{T} \cdot \epsilon_{\inp}$, where $T = \Theta_{k, \gamma}(1)$ is the maximum possible number of iterations of the template algorithm (see Lemma~\ref{lm:template}). Then it takes $\tilde{O}_{k, \gamma}(n^{2-\epsilon_{\inp}})$ time to  implement the template algorithm, under adjacency-matrix query access to $G$. Further, if the template algorithm does not return {\sc Failure}, then at the end of our implementation we have an oracle $\match_{M^{\out}}(.)$ for the matching $M^{\out}$ returned by it, with query time $\tilde{O}_{k, \gamma}(n^{1+\epsilon_{\out}})$.
\end{cor}

\begin{proof}
By Lemma~\ref{lm:sublinear:induction}, each iteration $t$ of the template algorithm can be implemented in time $\tilde{O}_{k, \gamma}(n^{2-\epsilon_{t-1}})  = \tilde{O}_{k, \gamma}(n^{2-\epsilon_{\inp}})$, since $\epsilon_{\inp} \leq \epsilon_{t-1}$. Thus, the total time taken to implement the template algorithm is at most  $\tilde{O}_{k, \gamma}(T \cdot n^{2-\epsilon_{\inp}}) = \tilde{O}_{k, \gamma}(n^{2-\epsilon_{\inp}})$.

Suppose that the template algorithm terminates at the end of iteration $t$, and returns a matching $M^{\out}$. Then, at the end of iteration $t$ of our sublinear implementation, the situation is as follows.
\begin{wrapper}
 $\sigma(t) = k$, and $\Lambda^{(t)} = \left\{ \lambda^{(t)}_0,  \lambda^{(t)}_1, \cdots, \lambda^{(t)}_{k}\right\}$,
where $\sigma\left( \lambda^{(t)}_j\right) = 2j$ for each $j \in [0, k]$ (see Invariant~\ref{new:inv:1}). The sequence of matchings in $\Lambda^{(t)}$ is nested (see Invariant~\ref{new:inv:nested}). Thus, from this sequence of nested matchings we can extract a set of at least $\left| M^{\left(\lambda^{(t)}_k\right)}\right|$ many node-disjoint length $(2k+1)$-augmenting paths w.r.t.~$M^{\inp}$ in $G$ (see Observation~\ref{new:obs:nested}). The template algorithm obtains the matching $M^{\out}$ by applying these augmenting paths to $M^{\inp}$. In our sublinear implementation of the template algorithm, however, we can access each matching $M 
\in \Lambda^{(t)}$ only via an oracle $\match_{M}(.)$, which has a query time of at most $\tilde{O}_{k, \gamma}(n^{1+\epsilon_{t}})$ (see~(\ref{eq:sublinear:reduction:1}) and Lemma~\ref{lm:sublinear:induction}). Furthermore, we can access the matching $M^{\inp}$ only via the oracle $\match_{M^{\inp}}(.)$, which also has a query time of at most $\tilde{O}_{k, \gamma}(n^{1+\epsilon_{\inp}}) = \tilde{O}_{k, \gamma}(n^{1+\epsilon_{t}})$.
\end{wrapper}

We now show how to answer a query to the oracle $\match_{M^{\out}}(v)$. The key observation is this:
\begin{wrapper}
 Let $E^* := (M^{\inp} \cap E_H) \bigcup_{M \in \Lambda^{(t)}} M$ (see the discussion on layered subgraphs in Section~\ref{sec:template:describe}). Then the  graph $G^* = (V, E^*)$ consists of a collection of node-disjoint alternating paths w.r.t.~$M^{\inp}$. 
We say that a path in $G^*$ is {\em complete} iff it has one endpoint at layer $0$ and the other endpoint at layer $(2k+1)$. Now, a node $v \in V$ is matched in $M^{\out}$ iff: either $v \in V(M^{\inp})$, or $v \notin V(M^{\inp})$ and $v$ is the starting/end point of a complete path in $G^*$.
\end{wrapper}

Using this observation, we now describe how to answer queries of the form: ``Is $\match_{M^{\out}}(v) = \perp$ for a given node $v \in V$?''. To answer such a query, we apply the  procedure below. 

If $\match_{M^{\inp}}(v) \neq \perp$,  then we return that $\match_{M^{\out}}(v) \neq \perp$. Else if $\match_{M^{\inp}}(v) = \perp$ and $\ell(v) \notin \{0, 2k+1\}$, then we return that $\match_{M^{\out}}(v) = \perp$. Finally, if $\match_{M^{\inp}}(v) = \perp$ and w.l.o.g.~$\ell(v) = 0$, then we perform the following steps.
\begin{itemize}
\item $v_0 \leftarrow v$.
\item {\bf For} $i = 1$ to $2k+1$:
\begin{itemize}
\item {\bf If} $i$ is odd, {\bf then}
 $v_i \leftarrow \mate_{M^{\left( \lambda^{(t)}_{(i-1)/2}\right)}}(v_{i-1})$.
\item {\bf Else if} $i$ is even, {\bf then}
$v_i \leftarrow \mate_{M^{\inp}}(v_{i-1})$.
\item {\bf If} $v_i = \perp$, {\bf then} return that $\match_{M^{\out}}(v) = \perp$.
\end{itemize}
\item Return that $\match_{M^{\out}}(v) \neq \perp$.
\end{itemize}

It is easy to verify that the above procedure correctly returns whether or not $\match_{M^{\out}}(v) = \perp$. We can extend this procedure in a natural manner, which would also allow us to answer the query $\match_{M^{\out}}(v)$. To summarize,  we can answer a query $\match_{M^{\out}}(v)$ by making at most one call to each of the oracles $\match_{M}(.)$, for $M \in \Lambda^{(t)}$, and at most $\Theta(k)$ calls to the oracle $\match_{M^{\inp}}(.)$. Each of these oracle calls take at most $\tilde{O}_{k, \gamma}(n^{1+\epsilon_t})$ time, as $\epsilon_{t'} \leq \epsilon_t$ for all $t' \in [1, t]$. Since $\left| \Lambda^{(t)} \right| = k$, the oracle $\match_{M^{\out}}(.)$ has a query time of $\tilde{O}_{k, \gamma}(k \cdot n^{1+\epsilon_t}) = \tilde{O}_{k, \gamma}(n^{1+\epsilon_{t}}) = \tilde{O}_{k, \gamma}(n^{1+\epsilon_{\out}})$, where the last equality holds since $\epsilon_t \leq \epsilon_{T} = \epsilon_{\out}$. This concludes the proof.
\end{proof}

\subsubsection*{Proof of Lemma~\ref{lm:sublinear:induction}}
\label{sec:proof:lm:sublinear:induction}

We prove the lemma by induction on $t$. 

\paragraph{Base case $(t = 0)$:} 

\medskip
\noindent 
We already have the oracle $\match_{M^{(0)}}(.)$ with query time $\tilde{O}_{k, \gamma}(n^{1+\epsilon_{0}})$, since $\epsilon_0 = \epsilon_{\inp}$ and $M^{(0)} = M^{\inp}$. We set $\sigma(t) \leftarrow \perp$, $\stack(t) \leftarrow -1$ and $\Lambda^{(0)} \leftarrow \emptyset$. We now claim that we already have the oracle $\alive_{0}(.)$. This is because a node $v \in V$ is alive just before the start of iteration $1$ if and only if $v \in V_H$.  Furthermore, given a query $\alive_{0}(v)$, we can determine whether $v$ is in $V_H$ by checking the value of $\ell(v)$,  setting $u \leftarrow \mate_{M^{(0)}}(v)$, and then checking the value of $\ell(u)$ if $u \neq \perp$. Thus,  answering a query to the oracle $\alive_0(.)$ takes $\tilde{O}_{k, \gamma}(n^{1+\epsilon_0})$ time. So, we can implement iteration $0$ in $O(1)$ time, and Lemma~\ref{lm:sublinear:induction} holds for $t = 0$.

\paragraph{Inductive case $(t \geq 1)$:} 

\medskip\noindent 
We assume that Lemma~\ref{lm:sublinear:induction} holds for all $t' \in [0, t-1]$, and that we have access to the data structures constructed during all these past iterations. We now focus on implementing the current iteration $t$ (see Section~\ref{sec:template:describe}) under adjacency-matrix query access to $G$. Let $i = \stack(t-1)$. If $i = k$, then the algorithm would terminate after  iteration $(t-1)$. Henceforth, we assume that $i \in [-1, k-1]$. In the current iteration $t$, the template algorithm wants to first compute a matching $M^{(t)}$ by calling the subroutine $\largematch(C_{2i+2} \cup A_{2i+3}, \delta_{i+1})$. We first show that we can efficiently query whether or not a given node in $V$ belongs to the set $C_{2i+2} \cup A_{2i+3}$. Subsequently, we split up our implementation of iteration $t$ into two steps, as described below.

\begin{claim}
\label{cl:query:candidate}
Given any node $v \in V$, we can determine if $v \in C_{2i+2}$ in $\tilde{O}_{k, \gamma}(n^{1+\epsilon_{t-1}})$ time.
\end{claim}

\begin{proof}
We first  check the value of $\ell(v)$ and  call  $\alive_{t-1}(v)$. Now, we consider the following cases. 
\begin{enumerate}[(i)]
\item $\ell(v) \neq 2i+2$. Here, we return that $v \notin C_{2i+2}$.
\item $\ell(v) = 2i+2$ and $\alive_{t-1}(v) = \text{{\sc No}}$. Here, we also return that $v \notin C_{2i+2}$.
\item $\ell(v) = 2i+2$, $\alive_{t-1}(v) = \text{{\sc Yes}}$, and $i = -1$. Here, we return that $v \in C_{2i+2}$.
\item $\ell(v) = 2i+2$, $\alive_{t-1}(v) = \text{{\sc Yes}}$ and $i \geq 0$. Here, we first set  $u_v \leftarrow \mate_{M^{\inp}}(v)$, which takes $\tilde{O}_{k, \gamma}(n^{1+\epsilon_{\inp}}) = \tilde{O}_{k, \gamma}(n^{1+\epsilon_{t-1}})$ time. Next, we call $\match_{M^{\left(\lambda^{(t-1)}_i\right)}}(u_v)$,\thatchaphol{Notation: can we change from $\match_{M^{\left(\lambda^{(t-1)}_i\right)}}(u_v)$ to $\match_{M^{(\lambda_{i}(t))}}(u_v)$ for readability?} which also takes at most $\tilde{O}_{k, \gamma}(n^{1+\epsilon_{t-1}})$ time. Finally, we return that $v \in C_{2i+2}$ iff $\match_{M^{\left(\lambda^{(t-1)}_i\right)}}(u_v) \neq \perp$.
\end{enumerate}
The correctness of the above procedure follows from the definition of the set $C_{2i+2}$. Furthermore, the preceding discussion implies that this procedure overall takes at most $\tilde{O}_{k, \gamma}(n^{1+\epsilon_{t-1}})$ time.
\end{proof}

\begin{cor}
\label{cor:query:candidate}
Given any  $v \in V$, we can determine if $v \in C_{2i+2} \cup A_{2i+3}$ in $\tilde{O}_{k, \gamma}(n^{1+\epsilon_{t-1}})$ time.
\end{cor}

\begin{proof}
We can determine if $v \in A_{2i+3}$ by checking the value of $\ell(v)$ and making a query $\alive_{t-1}(v)$, which takes $\tilde{O}_{k, \gamma}(n^{1+\epsilon_{t-1}})$ time. The corollary now follows from Claim~\ref{cl:query:candidate}.
\end{proof}

\paragraph{Step I: Constructing the oracle $\match_{M^{(t)}}(.)$.}  Armed with Corollary~\ref{cor:query:candidate}, we  mimic the call to  $\largematch(C_{2i+2} \cup A_{2i+3}, \delta_{i+1})$ in the template algorithm, by invoking Theorem~\ref{thm:matching oracle} with $A = C_{2i+2} \cup A_{2i+3}$,  $\delta_{\inp} = \delta_{i+1}$, $\epsilon = 2 \cdot \epsilon_{t-1}$ and $t_A = \tilde{O}_{k, \gamma}(n^{1+\epsilon_{t-1}})$.\footnote{The reader should keep in mind that in the current section (Section~\ref{sec:sublinear:main}), we are using the symbol $A$ to denote the set of alive nodes across all the layers. This is different from the way the symbol $A$ is being used in the statement of Theorem~\ref{thm:matching oracle}, where it refers to any arbitrary subset of nodes.} If Theorem~\ref{thm:matching oracle} returns $\perp$, then we set $M^{(t)} := \emptyset$, and  the  trivial oracle $\match_{M^{(t)}}(.)$ has $O(1) = \tilde{O}_{k, \gamma}(n^{1+\epsilon_t})$ query time. Otherwise, Theorem~\ref{thm:matching oracle} returns  an oracle $\match_{M}(.)$ for a matching $M$, and we set $M^{(t)} := M$. By~(\ref{eq:sublinear:reduction:1}) and  Theorem~\ref{thm:matching oracle}, this oracle $\match_{M^{(t)}}(.)$ has query time: 
\begin{eqnarray*}
\tilde{O}\left( \frac{(t_A + n) \cdot n^{4 \epsilon}}{\poly(\delta_{i+1})}\right) = \tilde{O}_{k, \gamma}\left( (t_A + n) \cdot n^{4 \epsilon}\right) = \tilde{O}_{k, \gamma}\left( n^{1+\epsilon_{t-1} + 4 \epsilon} \right) = \tilde{O}_{k, \gamma}\left(n^{1+\epsilon_t} \right).
\end{eqnarray*}
Finally, from~(\ref{eq:sublinear:reduction:1}) and Theorem~\ref{thm:matching oracle}, we infer that overall Step I takes time:
\begin{eqnarray*}
\tilde{O}\left(\frac{(t_A + n) \cdot (n^{1-\epsilon} + n^{4 \epsilon})}{\poly(\delta_{i+1})} \right) & = & \tilde{O}_{k, \gamma}\left( (t_A+n) \cdot (n^{1-\epsilon} + n^{4\epsilon}) \right) \\
& = & \tilde{O}_{k, \gamma}((t_A+n) \cdot n^{1-\epsilon})  \\
& = & \tilde{O}_{k, \gamma}\left( n^{2+\epsilon_{t-1} - \epsilon}\right) \\
& = & \tilde{O}_{k,\gamma}\left( n^{2-\epsilon_{t-1}}\right).
\end{eqnarray*}
In the above derivation, the second equality holds since  $\epsilon =2  \epsilon_{t-1} \leq 9^T  \epsilon_{\inp} \leq 1/5$ (see~(\ref{eq:sublinear:reduction:1}) and Definition~\ref{def:parameters}), whereas the third equality holds since $t_A = \tilde{O}_{k, \gamma}(n^{1+\epsilon_{t-1}})$.

\paragraph{Step II: Determining  $\sigma(t), \stack(t)$,  $\Lambda^{(t)}$, and the oracle $\alive_{(t)}(.)$.} We set $\sigma(t) \leftarrow 2i+2$. 
We now fork into one of the following three cases.
\begin{description}
\item[Case (a)] In Step I, the invocation of Theorem~\ref{thm:matching oracle} returned an oracle $\match_{M}(.)$ for a matching $M$, and we set $M^{(t)} := M$. This will be referred to as a forwarding iteration at layer $2i+2$. In this case, we set $\Lambda^{(t)} \leftarrow \Lambda^{(t-1)} \cup \{ t \}$ and $\stack(t) \leftarrow \stack(t-1) + 1$. Now, we observe that the set of alive nodes does not change during such a forwarding iteration, and so we already have the oracle $\alive_{t}(.)$, because $\alive_t(v) = \alive_{(t-1)}(v)$ for all $v \in V$. Accordingly, the oracle $\alive_t(.)$ has query time $\tilde{O}_{k, \gamma}(n^{1+\epsilon_{t-1}}) = \tilde{O}_{k, \gamma}(n^{1+\epsilon_t})$. 
\item[Case (b):] 
 In Step I, the invocation of Theorem~\ref{thm:matching oracle} returned $\perp$, and $i = -1$.  Here,  the algorithm terminates and returns {\sc Failure}.
\item[Case (c):] In Step I, the invocation of Theorem~\ref{thm:matching oracle} returned $\perp$, and $i \geq 0$. This will be referred to as a backtracking iteration at layer $2i$. In this case, we set $\Lambda^{(t)} \leftarrow \Lambda^{(t-1)} \setminus \{ \lambda^{(t-1)}_i \}$ and $\stack(t) \leftarrow \stack(t-1) - 1$. 
  Now, we observe that due to iteration $t$, only the nodes in $C_{2i+2}$ and their matched neighbors under $M^{\inp}$ (who are at layer $2i+1$), change their status from alive to dead. The status of every other node remains unchanged. Thus,  we can answer a query $\alive_{t}(v)$, in $\tilde{O}_{k,\gamma}(n^{1+\epsilon_t})$ time, as follows.

We first check the value of $\ell(v)$,   query $\alive_{t-1}(v)$ and $\match_{M^{\inp}}(v)$, and determine whether or not $v \in C_{2i+2}$ by invoking Claim~\ref{cl:query:candidate}. Overall, this takes $\tilde{O}_{k, \gamma}(n^{1+\epsilon_{t-1}}) + \tilde{O}_{k, \gamma}(n^{1+\epsilon_{\inp}}) = \tilde{O}_{k, \gamma}(n^{1+\epsilon_t})$ time. 
Next, we consider three cases. 
\begin{enumerate}[(i)]
\item $\ell(v) \notin \{2i+1, 2i+2\}$. Here, we return  $\alive_t(v) = \alive_{(t-1)}(v)$. 
\item $\ell(v) = 2i+2$. Here, if $v \in C_{2i+2}$ then we return $\alive_t(v) = \text{{\sc No}}$;  otherwise we return $\alive_t(v) = \alive_{t-1}(v)$. 
\item $\ell(v) = 2i+1$. Here, we set $u_v \leftarrow \mate_{M^{\inp}}(v)$. Now, if $u_v \in C_{2i+2}$ then  we return $\alive_t(v) = \text{{\sc No}}$; otherwise we return  $\alive_t(v) = \alive_{t-1}(v)$.
\end{enumerate}
\end{description}

To summarize, Step I takes $\tilde{O}_{k, \gamma}(n^{2-\epsilon_{t-1}})$ time, whereas Step II takes only $O(k)={O}_{k,\gamma}(1)$ time. Furthermore, at the end of Step II we have  all the desired data structures for iteration $t$, and both the oracles $\match_{M^{(t)}}(.)$ and $\alive_t(.)$ have a query time of at most $\tilde{O}(n^{1+\epsilon_t})$. Finally, Theorem~\ref{thm:matching oracle} ensures that whp, the way we decide whether we are in case (a), case (b) or case (c) is consistent with the choice  made by the template algorithm in the same scenario (see the discussion on ``implementing iteration $t$'' in Section~\ref{sec:template:describe}, and how the subroutine $\largematch(S, \delta)$ is defined in the second paragraph of Section~\ref{sec:algo:description}). This concludes the proof of Lemma~\ref{lm:sublinear:induction}.

%% file: 6-sublinear.tex
\section{$(1, \epsilon n)$-Approximate Matching Oracle}
\label{sec:sublinear:main}


In this section, starting from an empty matching, we repeatedly apply Theorem~\ref{thm:boost} to obtain our main results in the sublinear setting. They are summarized in the theorem and the corollary below, which are restatements of  Theorem~\ref{thm:main sublinear}.

\begin{thm}
\label{th:sublinear:main}
Let $\gamma, \epsilon'' \in (0, 1)$ be any two small constants, and let $G$ be the input graph with $n$ nodes which we can access via adjacency-matrix queries. Then for a sufficiently small constant $\epsilon' \in (0, \epsilon'')$, there exists  an algorithm which:
In $\tilde{O}_{\gamma}(n^{2-\epsilon'})$ time, returns an oracle $\match_{M}(.)$ for a $(1, 3\gamma n)$-approximate matching $M$ in $G$, where the oracle $\match_{M}(.)$ has $\tilde{O}_{\gamma}(n^{1+\epsilon''})$ query time.
\end{thm}
\begin{cor}
\label{cor:sublinear:main}
Given adjacency-matrix query access to an $n$-node graph $G$ and any constant $\gamma \in (0, 1)$, in $\tilde{O}_{\gamma}(n^{2-\epsilon})$ time we can return a $(1, 4\gamma n)$-approximation to the value of $\mu(G)$, whp. Here, $\epsilon \in (0, 1)$ is a sufficiently small constant depending on $\gamma$.
\end{cor}

\begin{proof}
First, we apply Theorem~\ref{th:sublinear:main}, with $\epsilon = \epsilon'$, to get the oracle $\match_{M}(.)$ in $\tilde{O}_{\gamma}(n^{2-\epsilon})$ time. Note that $M$ is a $(1, 3\gamma n)$-approximate matching in $G = (V, E)$. Using Chernoff bound, we  now  compute a $(1, \gamma n)$-approximate estimate $\hat{\mu}$ of $|M|$ by sampling, uniformly at random, a set $S$ of $\tilde{O}_{\gamma}(1)$ nodes from $V$ and querying $\match_{M}(v)$ for each node $v \in S$. This takes  $\tilde{O}_{\gamma}(n^{1+\epsilon''}) = \tilde{O}_{\gamma}(n^{2-\epsilon})$ time. The last inequality holds since $\epsilon = \epsilon'$ and $\epsilon''$ are chosen to be sufficiently small, so that $1+\epsilon'' \leq 2-\epsilon$. It is now easy to observe that $\hat{\mu}$ is a $(1, 4\gamma n)$-approximation to the value of $\mu(G)$.
\end{proof}

\subsection*{Proof of Theorem~\ref{th:sublinear:main}} Algorithm~\ref{alg:sublinear} contains the relevant pseudocode. We slightly abuse the notation in step 2-(b) of Algorithm~\ref{alg:sublinear}, when we write $Z = (M^{\out}, \epsilon_{\out})$. Here, we essentially mean that $\algo(G, M^{\inp}, i, \gamma^2, \epsilon_{\inp})$ returns the oracle $\match_{M^{\out}}(.)$ with query time $\tilde{O}_{\gamma}(n^{1+\epsilon_{\out}})$. Similarly, in step 2-(b), when we write $M^{\inp} \leftarrow M^{\out}$, this means that henceforth we will refer to the oracle $\match_{M^{\out}}(.)$ as $\match_{M^{\inp}}(.)$. 

The idea behind Algorithm~\ref{alg:sublinear} is simple and intuitive. We start by initializing $M^{\inp} \leftarrow \emptyset$, $k \leftarrow \lceil 1/\gamma \rceil$ and $\epsilon_{\inp} \leftarrow \epsilon'$, where $\epsilon' \in (0, 1)$ is a sufficiently small constant. At this point, we trivially have the oracle $\match_{M^{\inp}}(.)$ with query time $\tilde{O}_{\gamma}(n^{1+\epsilon_{\inp}})$. The algorithm now runs in {\em rounds}.  In each round, it repeatedly tries to augment the matching $M^{\inp}$ along small-length augmenting paths, by successively calling $\algo(G, M^{\inp}, i, \gamma^2, \epsilon_{\inp})$ for $i \in [0, k]$. Whenever a call to $\algo(\cdot)$ succeeds, the algorithm feeds its output into the next call to $\algo(\cdot)$. The algorithm terminates whenever it encounters a round where every call to $\algo(\cdot)$ returns {\sc Failure}.

\begin{algorithm}
Choose $\epsilon' \in (0, 1)$ to be a sufficiently small constant.

$\epsilon_{\inp} \leftarrow \epsilon'$, $k \leftarrow \lceil 1/\gamma \rceil$,  $M^{\inp} \leftarrow \emptyset$.

$\tau \leftarrow \text{{\sc True}}$.\\
\ 
\textbf{While $\tau = \text{{\sc True}}$:} \qquad // Start of a new round 
\begin{enumerate}
\item $\tau \leftarrow \text{{\sc False}}$.
\item {\bf For} $i = 0$ to $k$:
\begin{enumerate}
\item $Z \leftarrow \algo(G, M^{\inp}, i, \gamma^2, \epsilon_{\inp})$. \qquad // See Theorem~\ref{thm:boost}
\item {\bf If} $Z \neq \text{{\sc Failure}}$, {\bf then} 
\begin{itemize}
\item Suppose that $Z = (M^{\out}, \epsilon^{\out})$. 
\item $M^{\inp} \leftarrow M^{\out}$, $\epsilon_{\inp} \leftarrow \epsilon_{\out}$.
\item $\tau \leftarrow \text{{\sc True}}$.
\end{itemize}
\end{enumerate}
\end{enumerate}
$M \leftarrow M_{\inp}$, $\epsilon'' \leftarrow \epsilon_{out}$.

\textbf{Return the oracle 
$\match_{M}(\cdot)$, which has query time $\tilde{O}_{\gamma}(n^{1+\epsilon''})$.}

\caption{Near-optimal-matching-oracle $(G = (V, E), \gamma)$.\label{alg:sublinear}}

\end{algorithm}

\begin{claim}
\label{cl:bound:round}
Algorithm~\ref{alg:sublinear} runs for $\tilde{O}_{\gamma}(1)$ rounds, and makes  $\tilde{O}_{\gamma}(1)$ calls to $\algo(\cdot)$.
\end{claim}

\begin{proof}
Say that a given round of Algorithm~\ref{alg:sublinear} is {\em successful} iff during that round: for some $i \in [0, k]$, the call to $\algo(G, M^{\inp}, i, \gamma^2, \epsilon_{\inp})$ succeeded (see Theorem~\ref{thm:boost} and step 2-(a) of Algorithm~\ref{alg:sublinear}). By Theorem~\ref{thm:boost}, each time a call to $\algo(G, M^{\inp}, i, \gamma^2, \epsilon_{\inp})$ succeeds, it increases the size of the matching $M^{\inp}$ (see step 2-(b) of Algorithm~\ref{alg:sublinear}) by at least $\Theta_{\gamma}(1) \cdot n$. Since $\mu(G) \leq n$, such an event can occur at most $\Theta_{\gamma}(1)$ times. Finally,  each round of Algorithm~\ref{alg:sublinear} consists of $(k+1) = \Theta_{\gamma}(1)$ calls to $\algo(\cdot)$, and all but the last round is successful. This implies the claim.
\end{proof}

\begin{claim}
\label{cl:success}
Suppose that at the start of a given round of Algorithm~\ref{alg:sublinear}, there exists a collection of at least $\gamma^2 \cdot n$ many node-disjoint length $(2i+1)$-augmenting paths w.r.t.~$M^{\inp}$ in $G$, for some $i \in [0, k]$. Then whp, Algorithm~\ref{alg:sublinear} does {\em not} terminate at the end of the given round.
\end{claim}

\begin{proof}
If there exists some $j \in [0, i-1]$ such that the call to $\algo(G, M^{\inp}, j, \gamma^2,\epsilon_{\inp})$ succeeds during the given round, then it immediately implies the claim (since we would have $\tau = \text{{\sc True}}$ when the round ends and so the {\bf While} loop in Algorithm~\ref{alg:sublinear} will run for at least one more iteration). 

For the rest of the proof  assume that during the given round, for all $j \in [0, i-1]$ the call to $\algo(G, M^{\inp}, j, \gamma^2, \epsilon_{\inp})$ returns {\sc Failure}, and hence the matching $M^{\inp}$ does not change during iterations $j = 0$ to $i-1$ of the {\bf For} loop. Accordingly, at the start of the concerned iteration $i$ of the {\bf For} loop, the matching $M^{\inp}$ still admits a collection of at least $\gamma^2 \cdot n$ many node-disjoint length $(2i+1)$-augmenting paths in $G$. Thus, by Theorem~\ref{thm:boost},  the call to $\algo(G, M^{\inp}, i, \gamma^2, \epsilon_{inp})$ succeeds whp. So, it follows that Algorithm~\ref{alg:sublinear} does not end after the given round, whp.
\end{proof}

\begin{cor}
\label{cor:sublinear:approx:main} When Algorithm~\ref{alg:sublinear} terminates, whp $M^{\inp}$ is a $(1, 3\gamma n)$-approximate matching in $G$.
\end{cor}

\begin{proof}
Let $M^*$ be a maximum matching in $G$. By Claims \ref{cl:bound:round} and \ref{cl:success}, the following  holds whp when the algorithm terminates:
For all $i \in [0, k]$, there exists at most $\gamma^2 \cdot n$ many length-$(2i+1)$ augmenting paths in $M^{\inp} \cup M^*$.

As $k = \lceil 1/\gamma \rceil$, the augmenting paths in $M^{\inp} \cup M^*$ that are of length $\leq 2k+1$ contribute at most $(k+1) \cdot \gamma^2 n \leq 2 \gamma  n$  extra edges to $M^*$ compared to $M^{\inp}$. On the other hand, augmenting paths $M^{\inp} \cup M^*$ that are of length $> (2k+1)$ contribute at most $\frac{(k+2) - (k+1)}{k+1} \cdot |M^{\inp}| \leq \gamma \cdot |M^{\inp}| \leq \gamma n$ extra edges to $M^*$ compared to $M^{\inp}$. Thus, we get: $|M^*| \leq |M^{\inp}| + 3\gamma n$.
\end{proof}

Since Algorithm~\ref{alg:sublinear} makes only constantly many calls to $\algo(\cdot)$, we can choose  $\epsilon' > 0$ to be sufficiently small so as to guarantee that $0 < \epsilon'' \ll 1$ (see Claim~\ref{cl:bound:round} and Theorem~\ref{thm:boost}). Further, during the execution of Algorithm~\ref{alg:sublinear}, each call to $\algo(\cdot)$ takes $\tilde{O}_{\gamma}(n^{2-\epsilon_{\inp}}) = \tilde{O}_{\gamma}(n^{2-\epsilon'})$ time. Theorem~\ref{th:sublinear:main} now follows from Claim~\ref{cl:bound:round} and  Corollary~\ref{cor:sublinear:approx:main}.

%% file: 7-dynamic.tex
\section{Dynamic $(1+\eps)$-Approximate Matching Size}
\label{sec:dynamic}

We now prove our main result in the dynamic setting; as summarized in the theorem below. Note that Theorem~\ref{new:thm:main dynamic} is a restatement of Theorem~\ref{thm:main dynamic}.


\begin{thm}
\label{new:thm:main dynamic}There is a dynamic $(1+\eps)$-approximate matching \emph{size} algorithm with $m^{0.5-\Omega_{\epsilon}(1)}$ worst-case update time, where $m$ is the number of edges in the dynamic input graph $G = (V, E)$ with $n$ nodes. The algorithm is randomized and works against an adaptive adversary whp. Moreover, the algorithm maintains an oracle $\match_M(.)$ with query time $\tilde{O}(m^{0.5 + \eps'})$ (for a small constant $\eps' > 0$ which depends on $\eps$), where $M$ is a $(1+\epsilon)$-approximate maximum matching of $G$.
\end{thm}

To highlight the main idea behind the proof of Theorem~\ref{new:thm:main dynamic}, first we recall that using techniques presented in a series of papers~\cite{AssadiKL19,behnezhad2023dynamic,BehnezhadDH20,bhattacharya2023dynamic,kiss2022improving}, we can assume:  $\mu(G) = \Omega(n)$ throughout the sequence of updates. Accordingly, consider the following  dynamic matching size algorithm, which runs in phases, where each phase  lasts for $\epsilon n$ updates. At the start of a phase, we  compute a $(1+\epsilon)$-approximate estimate $\mu^*$ of $\mu(G)$, by invoking Corollary~\ref{cor:sublinear:main}, in $\tilde{O}_{\epsilon}(n^{2-\epsilon})$ time.  Sine $\mu(G) = \Omega(n)$, the value of  $\mu^*$ continues to remain a $(1+\Theta(\epsilon))$-approximate estimate of $\mu(G)$ throughout the duration of the phase. This already leads to an amortized update time of: $\tilde{O}_{\epsilon}(n^{2-\epsilon})/(\epsilon n) = \tilde{O}_{\epsilon}(n^{1-\epsilon})$, which is sublinear in $n$. We now show how to extend this idea to get an update time that is sublinear in $\sqrt{m}$, and how to answer queries in $m^{0.5+\epsilon'}$ time.

\subsection*{Proof of Theorem~\ref{new:thm:main dynamic}}
\label{sec:proof:thm:main dynamic}
For ease of exposition, we first focus on proving an amortized update time bound. We start by recalling a useful technique for sparsifying $G$, which  allows us to assume that $\mu(G) = \Omega(n)$.

\paragraph{Contractions:}  Consider a function $\phi : V \rightarrow V_{\phi}$ which maps every node in $V$ to some element in the set $V_{\phi}$. We say that $\phi$ is a {\em contraction} of $G$ iff $|V_{\phi}| \leq |V|$. Define the multiset of edges $E_{\phi} := \{ (u, v) \in E : \phi(u) \neq \phi(v) \}$, and consider the multigraph $G_{\phi} := (V_{\phi}, E_{\phi})$. From every matching in $G_{\phi}$, we can recover a matching in $G$ of  same size. Hence, we have: $\mu(G_{\phi}) \leq \mu(G)$.  

\medskip
The next theorem follows immediately from  past work on the maximum matching problem across a range of computational models~\cite{AssadiKL19,behnezhad2023dynamic,BehnezhadDH20,bhattacharya2023dynamic,kiss2022improving}.  For the sake of completeness, however, we outline the proof of \Cref{thm:contraction} in Appendix~\ref{app:sec:dynamic}.

\newcommand{\A}{\mathcal{A}}

\begin{lem}
\label{thm:contraction}
There exists a dynamic  algorithm $\A$ with $\tilde{O}(1)$ worst-case update time, which maintains: a set of $K = \tilde{O}(1)$ contractions $\{\phi_1, \ldots, \phi_K\}$ of $G$, the corresponding graphs $\{G_{\Phi_1}, \ldots, G_{\Phi_K}\}$, and a subset  $I \subseteq [1, K]$. Throughout the sequence of updates (whp against an adaptive adversary) the algorithm ensures that: (i) $|V_{\phi_i}| = \Theta\left( \frac{\mu(G)}{\epsilon} \right)$ for all $i \in I$, and (ii) there is an index $i^* \in I$ such that $(1-\epsilon) \cdot \mu(G) \leq \mu(G_{\phi_{i^*}}) \leq  \mu(G)$.
\end{lem}

\paragraph{Description of our dynamic algorithm:} 
We maintain a $(2+\epsilon)$-approximate estimate $\hat{\mu}$ of $\mu(G)$, in $\tilde{O}(1)$ worst-case update time, using an existing deterministic dynamic  matching algorithm as a subroutine~\cite{bhattacharya2017deterministic}. We also use the algorithm $\A$, as  in Theorem~\ref{thm:contraction}, as a subroutine. Let $\epsilon_0 \in (0, 1)$ be a sufficiently small constant, depending on $\epsilon$. 

Our dynamic algorithm partitions the update sequence into {\em phases}. We now explain how the algorithm works during a given phase, which can be of two types.

\newcommand{\init}{\texttt{init}}

\paragraph{Type-I Phase:} At the start of a type-I phase, we have $\hat{\mu} \geq |E|^{0.5+\epsilon_0}$. Let $m_{\init}$ denote the value of $|E|$ at the start of the phase.  Then the phase will last for the next $\epsilon \cdot (m_{\init})^{0.5+\epsilon_0}$ updates. At the start of the phase, we call an existing static algorithm to compute a $(1+\epsilon)$-approximate maximum matching $M$ of $G$,  which takes $O_{\epsilon}(m_{\init})$ time~\cite{duan2014linear}. Define $\mu^* = |M|$. Throughout the phase, $\mu^*$ continues to remain a $(1+2\epsilon)$-approximate estimate of $\mu(G)$, and we continue to output the same value $\mu^*$.  This leads to an amortized update time of: $$\frac{{O}_{\epsilon}(m_{\init})}{\epsilon \cdot (m_{\init})^{0.5+\epsilon_{0}}} = {O}_{\epsilon}\left((m_{\init})^{0.5-\epsilon_{0}}\right) = m^{0.5 - \Omega_{\epsilon}(1)}.$$
The last equality holds since $|E| = m = \Theta(m_{\init})$ throughout the phase. We can ensure that throughout the phase, the algorithm explicitly maintains $M$, which remains a $(1+2\eps)$-approximate maximum matching of $G$. This gives us the matching oracle $\match_{M}(.)$, with constant query time.

\paragraph{Type-II Phase:} At the start of a type-II phase, we have $\hat{\mu} < |E|^{0.5+\epsilon_0}$. Let $\hat{\mu}_{\init}$ and $m_{\init}$ respectively denote the value of  $\hat{\mu}$ and $|E|$ at the start of the phase.   The phase will last for the next $\epsilon  \hat{\mu}_{\init}$ updates. Hence,  $\mu(G)$ can change by at most a multiplicative  $(1+\epsilon)$ factor during the phase.

At the start of the phase, for each $i \in I$, we find a $(1, 4\gamma)$-approximate estimate $\mu^*_i$ of $\mu(G_{\phi_i})$. We obtain  $\mu^*_i$ by invoking  Corollary~\ref{cor:sublinear:main} on $G_{\phi_i}$, with $\gamma = \epsilon^2$.\footnote{It is trivial to verify that Corollary~\ref{cor:sublinear:main} holds even when applied on a multigraph.} This takes time:
$$\tilde{O}\left(|V_{\phi_i}|^{2-\epsilon^*}\right) = \tilde{O}_{\epsilon}\left((\mu(G))^{2-\epsilon^*}\right) = \tilde{O}_{\epsilon}\left((\hat{\mu}_{\init})^{2-\epsilon^*}\right),$$  where $\epsilon^* \in (0, 1)$ is a sufficiently small constant depending on $\epsilon$. Since $|I| = \tilde{O}(1)$, overall we  spend $\tilde{O}_{\epsilon}\left((\hat{\mu}_{\init})^{2-\epsilon^*}\right)$ time to compute $\mu^*_i$ for all $i \in I$. Next, in $|I| = \tilde{O}(1)$ time, we find an index $j \in I$ which maximizes the value  $\mu^*_{j}$. From Theorem~\ref{thm:contraction}, it follows that: 

\begin{eqnarray}
(1-\epsilon) \cdot \mu(G) - \Theta(\gamma) \cdot \Theta\left( \frac{\mu(G)}{\epsilon}\right) \leq \mu^*_j \leq \mu(G). \label{eq:dynamic:amortized:1}
\end{eqnarray}
As $\gamma = \epsilon^2$, we infer that $\mu^*_j$ is a purely multiplicative  $(1+\Theta(\epsilon))$-approximate estimate of $\mu(G)$. We continue to output the same value $\mu^*_j$ throughout the phase, since we have already observed that during the phase   $\mu(G)$ changes by at most a multiplicative $(1+\epsilon)$ factor.

The phase lasts for $\epsilon \hat{\mu}_{\init}$ updates. Accordingly, this leads to an amortized update time of:
 $$\frac{\tilde{O}_{\epsilon}\left((\hat{\mu}_{\init})^{2-\epsilon^*} \right)}{\epsilon \hat{\mu}_{\init}} = \tilde{O}_{\epsilon} \left( (\hat{\mu}_{\init})^{1-\epsilon^*}\right) = \tilde{O}_{\epsilon}\left((m_{\init})^{0.5+\epsilon_0 - \epsilon^*}\right) =  m^{0.5 - \Omega_{\epsilon}(1)}.$$
The last equality holds because we can ensure that $\epsilon_0$ is sufficiently small compared to $\epsilon^*$ (which, in turn, depends on $\epsilon$), and since $|E| = m = \Theta(m_{\init})$ throughout the duration of the phase.

Because of~(\ref{eq:dynamic:amortized:1}), at the start of the phase we can construct the  oracle $\match_M(.)$  by invoking Theorem~\ref{th:sublinear:main} on the graph $G_{\phi_{j^*}}$. Over the $\epsilon \hat{\mu}_{\init}$ edge updates of the phase,  $M$ continues to remain a $(1+O(\epsilon))$-approximate maximum matching in $G$. Finally, to maintain the oracle under edge insertions/deletions during the phase, we simply ignore edge deletions and assume that if a vertex is matched by a deleted edge of $M$ then it is unmatched.

\paragraph{Improving the update time bound to worst-case:} Recall that $\hat{\mu}_{init}$ denotes the value of $\hat{\mu}$ at the beginning of a phase. As observed after the initialization of a phase the output maintained by the algorithm remains $(1+O(\epsilon))$-approximate for the next $O(\epsilon \cdot \hat{\mu}_{init})$ updates. Furthermore, the total computational work done by the algorithm in both types of phases is upper bounded by $\tilde{O}(\hat{\mu}_{init}^{2-\epsilon^*})$ for some small constant $\epsilon^* \in (0,1)$ depending on the parameters of the algorithm. Let $G_i$ stand for the state of the input graph at the beginning of phase $i$ and let $\calA(G_i)$ stand for the output of the previously described algorithm initialized on $G_i$. Let $\hat{\mu}_{init,i}$ stand for the value of $\hat{\mu}_{init}$ at the beginning of phase $i$. We will now describe the behaviour of the worst-case update time algorithm.

The improved algorithm similarly initializes it's output to be $\calA(G_1)$. Throughout the first three phases it does not alter it's output and during the first two phases it doesn't complete any background computation. During phase $i$ for $i>2$ the algorithm calculates $\calA(G_{i-2})$ distributing the work evenly throughout the phase. Note that by phase $i$ the algorithm has complete knowledge of $G_{i-2}$. At the end of the same phase it switches it's output to be $\calA(G_{i-2})$. 

As the algorithm only outputs a matching size estimate and an oracle and not an actual matching this switch is done in constant time. Computing $\calA(G_{i-2})$ takes time proportional to $\tilde{O}(\hat{\mu}_{init,i-2}^{2-\epsilon^*})$ and is distributed over $\epsilon \cdot \hat{\mu}_{init,i}$ updates. As during a phase $\mu(G)$ may change by at most a $1+O(\epsilon)$ multiplicative factor we must have that $\hat{\mu}_{init,i} = \Theta(\hat{\mu}_{init,i-2})$. The amortized implementation amortizes the work of computing $\calA(G_{i-2})$ over $\epsilon \cdot \hat{\mu}_{init,i-2}$ updates. This implies that the worst-case update time guarantee of the delayed rebuild based algorithm matches the amortized versions update time within a constant factor.

%% file: A-simple-oracle.tex
\section{Matching Oracles on Low Degree Graphs}

\label{sec:simple LCA}

In this section, we prove \Cref{thm:simple LCA}. 

\paragraph{Preliminaries on Randomized Greedy Matching.}
A greedy maximal matching in $G$ with respect to an edge permutation $\pi$, denoted by $M=GMM(G,\pi)$ is a maximal matching obtained by scanning through edges with ordering defined by $\pi$, and for each edge $e$, include the edge $e$ into the matching if both of its end point are not matched. The matching oracle $\match_{M}(v)$ of \Cref{thm:simple LCA} simply returns $\VO(v)$ where the vertex oracle $\VO$ and the edge oracle $\EO$ are defined in \cite{behnezhad2022time} follows. 

\begin{algorithm}
\begin{enumerate}
\item Let $e_{1}=(v,u_{1}),\dots,e_{k}=(v,u_{k})$ be the edges incident to $v$ where $\pi(e_{1})<\dots<\pi(e_{k})$.
\item \textbf{for }$i=1,\dots,k$: if $\EO(e_{i},u_{i},\pi)=\TRUE$, then \textbf{return }$(v,u_{i})$.
\item \textbf{return} $\bot$
\end{enumerate}
\caption{$\protect\VO(v,\pi)$ \label{alg:VO}}
\end{algorithm}

\begin{algorithm}
\begin{enumerate}
\item \textbf{if }$\EO(e,u,\pi)$ is computed, then return the computed answer.
\item Let $e_{1}=(u,w_{1}),\dots,e_{k}=(u,w_{k})$ be the edges incident to $u$ where $\pi(e_{i})<\pi(e)$ and $\pi(e_{1})<\dots<\pi(e_{k})$.
\item \textbf{for }$i=1,\dots,k$: if $\EO(e_{i},w_{i},\pi)=\TRUE$, then \textbf{return }$\FALSE$.
\item \textbf{return} $\TRUE$.
\end{enumerate}
\caption{$\protect\EO(e,u,\pi)$ \label{alg:EO}}
\end{algorithm}

Let $T(v,\pi)$ denote the number of recursive calls $\EO(\cdot,\cdot,\pi)$ over the course of answering $\VO(v,\pi)$. The main theorem of \cite{behnezhad2022time} is as follows.
\begin{lem}
[Theorem 3.5 of \cite{behnezhad2022time}]Let $v$ be a random vertex and $\pi$ be a random permutation over edges, independent from $v$. 
\end{lem}

\[
\E_{v,\pi}[T(v,\pi)]=\alpha\triangleq O(d\log n).
\]

Given access to adjacency list, we can execute $\VO(v,\pi)$ using $O(T(v,\pi)\Delta)$ time straightforwardly. But Behnezhad \cite{behnezhad2022time} also showed that we can think of $T(v,\pi)$ as the running time, given access to the adjacency lists:
\begin{lem}
[Lemma 4.1 of \cite{behnezhad2022time}]\label{lem:recurse in one query}Let $v$ be an arbitrary vertex in a graph $G=(V,E)$. There is an algorithm that draws a random permutation $\pi$ over $E$, and determines whether $v$ is matched in $GMM(G,\pi)$ in time $\Otil(T(v,\pi)+1)$ having query access to the adjacency lists. The algorithm succeeds w.h.p. 
\end{lem}

\paragraph{Basic Properties of Randomized Greedy Matching.}
Recall that $\dbar$ is the given parameter where $\dbar\ge d$. We set the \emph{threshold} $\ell=\Theta(\dbar\log(n)/\epsilon)$ such that $\ell\ge\alpha\cdot\frac{8}{\epsilon}$. We have by Markov's inequality that
\begin{equation}
\Pr_{v,\pi}[T(v,\pi)>\ell]\le\epsilon/8.\label{eq:1}
\end{equation}
For any edge permutation $\pi$, let $f(\pi)=\Pr_{v\sim V}[T(v,\pi)>\ell]$ measure the fraction of vertices such that randomized greedy matching w.r.t.~$\pi$ makes many recursive calls exceeding the threshold $\ell$. We say that $\pi\in\Pi$ is \emph{great} if $f(\pi)\le\epsilon/2$. Observe that 
\begin{equation}
\Pr_{\pi}[\pi\text{ is great}]\ge1/4\label{eq:great perm}
\end{equation}
Otherwise, $\Pr_{v,\pi}[T(v,\pi)>\ell]\ge\Pr_{v,\pi}[T(v,\pi)>\ell\mid\pi\text{ is not great}]\Pr_{\pi}[\pi\text{ is not great}]>(\epsilon/2)\cdot(1/4)$ which contradicts \Cref{eq:1}. We also say that $\pi\in\Pi$ is \emph{good} if $f(\pi)\le\epsilon$, otherwise we say that it is \emph{bad}. 

\begin{algorithm}
\begin{enumerate}
\item \label{enu:testperm sample}Sample $r=1000\log(n)/\epsilon$ vertices independently: $v_{1},\dots,v_{r}$. 
\item \label{enu:testperm count}Let $X=|i\mid\{T(v_{i},\pi)>\ell\}|$ and $\tilde{f}=X/r$. 
\item If $\tilde{f}\le\frac{3}{4}\epsilon$, return ``yes''. Otherwise, return ``no''. 
\end{enumerate}
\caption{$\protect\testperm(\pi)$ \label{alg:testperm}}
\end{algorithm}

A simple procedure in \Cref{alg:testperm} accepts a great permutation and rejects a bad permutation with high probability. 
\begin{lem}
\label{lem:testperm}If $\pi$ is great, then $\testperm(\pi)$ returns ``yes'' with high probability. If $\pi$ is bad, then $\testperm(\pi)$ returns ``no'' with high probability. 
\end{lem}

\begin{proof}
If $\pi$ is great but $\testperm(\pi)$ returns ``no'', then we have $f(\pi)\le\epsilon/2$ but $\tilde{f}>3\epsilon/4$. Since $\E[X]=\epsilon\cdot f(\pi)$ and $X=\epsilon\cdot\tilde{f}$, we have
\[
X-\E[X]>r\epsilon/4.
\]
Applying Chernoff bound \Cref{prop:chernoff} with $t=r\epsilon/4$ and $\overline{\mu}=r\epsilon/2\ge\E[X]$, we have
\[
\Pr[X-\E[X]>r\epsilon/4]\le\exp(-\frac{(r\epsilon/4)^{2}}{3\overline{\mu}})\le\exp(-\frac{r\epsilon}{24})\le1/n^{10}.
\]

If $\pi$ is bad but $\testperm(\pi)$ returns ``yes'', then we have $f(\pi)\ge\epsilon$ but $\tilde{f}<3\epsilon/4$. This means that $X<\frac{3}{4}\E[X]$. Applying the standard Chernoff bound, i.e.,$\Pr[X<(1-\delta)\E[X]]\le\exp(-\frac{\delta^{2}\E[X]}{2})$ for $\delta\in[0,1]$, we have
\[
\Pr[X<\frac{3}{4}\E[X]]\le\exp(-\frac{(\frac{1}{4})^{2}\E[X]}{2})\le\exp(-\frac{r\epsilon}{32})\le1/n^{10}.
\]
\end{proof}
Now, we are ready to prove \Cref{thm:simple LCA}. 

\paragraph{Preprocessing.}
The preprocessing algorithm is as follows. First, independently sample $O(\log n)$ random edge-permutations $\pi_{1},\dots,\pi_{O(\log n)}$. For any $i$, if $\testperm(\pi_{i})$ returns ``yes'', then set $\pi^{*}\gets\pi_{i}$. 

We claim that $\pi^{*}$ is good w.h.p. Recall that each $\pi_{i}$ is great with probability at least $1/4$ by \Cref{eq:great perm}. So w.h.p. one of the permutation $\pi_{i}$ is great and so, by \Cref{lem:testperm}, $\testperm(\pi_{i})$ must return ``yes'' w.h.p. Moreover, also by \Cref{lem:testperm}, the returned permutation $\pi^{*}$is not bad w.h.p. That is, $\pi^{*}$ is good.

\paragraph{Query.}
Given a vertex $v$, we simply execute $\VO(v,\pi^{*})$ except that we return $\bot$ if it makes more than $\ell$ recursive calls. If $\VO(v,\pi^{*})=(v,v')$ returns a matched edge, to make sure that the answers on $v$ and $v'$ are consistent, we also call $\VO(v',\pi^{*})$. If it turns out that $\VO(v',\pi^{*})$ makes more than $\ell$ recursive calls, then we return $\bot$. Otherwise, $\VO(v',\pi^{*})$ must also return $(v,v')$ and then we return $(v,v')$. 

By construction, we obtain a matching oracle whose answers are consistent w.r.t. some fixed matching $M$. If $\ell=\infty$, then $M$ would be a normal randomized greedy maximal matching of size $|M|\ge\mu(G)/2$. This might not be true in reality as we set $\ell=\Theta(d\log(n)/\epsilon)$. But since $\pi^{*}$ is good, i.e., $f(\pi^{*})=\Pr_{v\sim V}[T(v,\pi)>\ell]\le\epsilon$. So 
\[
|M|\ge\mu(G)/2-\epsilon n.
\]
This completes the correctness of \Cref{thm:simple LCA}. It remains to analyze the running time.

\paragraph{Time analysis.}
Step \Cref{enu:testperm sample} of $\testperm(\cdot)$ takes $O(r)$ time as we just sample $r$ vertices in $G$. Step \Cref{enu:testperm count} takes time $r\cdot\Otil(\ell)$ time where the factor $\Otil(\ell)$ is by \Cref{lem:recurse in one query}. Since $r=\Theta(\log(n)/\epsilon)$ and $\ell=\Theta(\dbar\log(n)/\epsilon)$, each $\testperm$ takes $\Otil(\dbar/\epsilon^{2})$. We call $\testperm$ $O(\log n)$ times. So the total preprocessing time is $\Otil(d/\epsilon^{2})$. For the query time, we run $\VO$ and makes $O(\ell)$ recursive calls. Therefore, the total query time is $\Otil(\ell)=\Otil(d/\epsilon)$ time by \Cref{lem:recurse in one query}.

%% file: B-vertex-sparsification.tex
\section{Vertex Reduction for Dynamic Matching: Proof of \Cref{thm:contraction}}
\label{app:sec:dynamic}

\label{app:lm:dynamic:contraction}

This algorithm description is analogous to the algorithm in \cite{kiss2022improving} which extends previous algorithms from \cite{AssadiKL19,behnezhad2023dynamic,BehnezhadDH20,bhattacharya2023dynamic} to function against an adaptive adversary. Assume we are given $G = (V,E)$. Using a $\tilde{O}(1)$ worst-case update time deterministic algorithm from literature we maintain an $\alpha = O(1)$-approximate estimate $\hat{\mu}$ of $\mu(G)$. We make $\tilde{O}(1)$ guesses of $\mu(G)$, $1,\alpha,\alpha^2,\dots,\alpha^{k}$. For $\mu(G)$ guess $\alpha^i$ we will define $T = \frac{\ln(n) \cdot 512}{\eps^2} = \tilde{O}(1)$ contractions of $V$ $\phi^i_j : j \in [T]$ and contracted graphs $G_{\phi^i_j}$. If $\hat{\mu} \in [\alpha^{i},\alpha^{i+1})$ we define $\mu(G)$ guess $\alpha^i$ to be the accurate guess at the given time. If guess $\alpha^i$ is currently the accurate guess then the algorithm will maintain that i) for all $j \in [T]$ we have that $|V_{\phi^i_j}| = \Theta(\frac{\mu(G)}{\epsilon})$ and ii) there exists some $j \in [T]$ such that $(1-\epsilon) \cdot \mu(G) \leq \mu(G_{\phi^i_j}) \leq \mu(G)$.

We will now define how $\phi^i_j$ is generated. We define a set of vertices $|V_{\phi^i_j}| = \frac{8 \cdot \alpha^{i+1}}{\epsilon}$ and map vertices of $V$ to $V_{\phi^i_j}$ uniformly at random. Note that property $i)$ holds as for the accurate guess we must have that $\mu(G) = \Theta(\alpha^i)$. From the definition of vertex contractions for any contracted graph $G_{\phi^i_j}$ we must have that $\mu(G_{\phi^i_j}) \leq \mu(G)$. Therefore, it remains to show that if $\alpha^i$ is the currently accurate guess of $\mu(G)$ then there exists some $j \in [T]$ such that $(1-\epsilon) \cdot \mu(G) \leq \mu(G_{\phi^i_j})$.

Let's assume that $\alpha^i$ is the currently accurate guess of $\mu(G)$. Note that this implies that $\mu(G)/\alpha \leq \alpha^i \leq \mu(G) \cdot \alpha$. Let $S$ be an arbitrary subset of $V$ of size $2 \cdot \mu(G)$ representing the possible endpoints of a maximum matching. There can be $\binom{n}{2 \mu(G)} \leq n^{2 \cdot \mu(G)} \leq \exp(\ln(n) \cdot 2 \mu(G))$ such chooses of $S$.Fix some $j \in [T]$. For all $v$ vertices of $V_{\phi^i_j}$ define the event $X^v_j$ to be the indicator variable of the event $\phi^{-1}(v) \cap S \neq \emptyset$ and define $\bar{X}_j = \sum_{v \in V_{\phi^i_j}} X^v_j$.

\begin{claim}

\label{cl:app:contraction:neativeassociation}

For a fixed $j$ events $X^v_j$ are negatively associated random variables.

\end{claim}

The proof of Claim~\ref{cl:app:contraction:neativeassociation} appears in the appendix of \cite{kiss2022improving}. Let $\beta = \alpha^{i+1}/\mu(G) \in [1, \alpha]$. We first lower bound the expectation of $\hat{X}_j$:

\begin{eqnarray}
\mathbb{E}[X^j_i] & = & 1 - \Pr[S \cap V_i^j = \emptyset] \nonumber \\
& = & 1 - \left(1 - \frac{1}{|V_{\phi_j^i}|}\right)^{2\mu(G)} \nonumber \\
& = & 1 - \left(1 - \frac{\epsilon}{8 \cdot \mu(G) \cdot \beta} \right)^{2\mu(G)} \nonumber \\
& \geq & 1 - \exp\left(-\frac{\epsilon}{4 \cdot \beta} \right) \nonumber \\
& \geq & \frac{\eps \cdot (1 - \eps/(8 \cdot \beta))}{4 \cdot \beta} \label{eq:app:contractions:1}
\end{eqnarray}

Inequality~\ref{eq:app:contractions:1} holds for small values of $\eps$. Therefore, 
$$\mathbb{E}[\bar{X_j}] \geq |V_{\phi_i^j}| \cdot \frac{\eps \cdot (1 - \eps/(8 \cdot \beta))}{4 \cdot \beta} \geq 2\mu(G) \cdot (1 - \epsilon/(8\cdot \beta))$$. 

Now we apply Chernoff's bound on the sum of negatively associated random variables $\bar{X}_j$ to get that: 

\begin{eqnarray}
    \Pr\left(\bar{X}_j \leq 2\mu(G) \cdot (1 - \frac{\eps}{4 \cdot \beta})\right) & \leq & \Pr\left(\bar{X}_j \leq \E[\bar{X}_j] \cdot (1 - \frac{\eps}{8}\right)\nonumber \\
    & \leq & \exp\left(-\frac{\E[\bar{X}_j] \cdot \left(\frac{\eps}{8}\right)^2}{2}\right) \nonumber \\
    & \leq & \exp\left( -\frac{\mu(G) \cdot \eps^2}{64}\right) \nonumber
\end{eqnarray}

Recall that we construct $T = \frac{\ln(n) \cdot 512}{\eps^2}$ contracted $G_{\phi_i^j}$ for $\mu(G)$ guess $\alpha^i$.

$$ \Pr\left( \min_{j \in [T]} \bar{X}_j \leq 2 \cdot \mu(G) \cdot (1 - \frac{\eps}{4})\right) \leq \left(1 - \exp\left( -\frac{\mu(G) \cdot \eps^2}{64}\right)\right)^{T} \leq \exp(-16 \ln(n) \cdot \mu(G))$$

Further recall that $S$ may be selected at most $\exp(2\ln(n)\cdot \mu(G))$ different ways. Hence, taking a union bound over the possible choices of $S$ we can say that regardless how $S$ was chosen there is a vertex contraction $\phi^i_j$ such that $\bar{X}_j \geq 2 \cdot \mu(G) \cdot (1-\eps/4)$. Fix any maximum matching $M^*$ of $G$. By this argument we know that with high probability there must be some $j \in [T]$ such that $\phi^i_j(V[M^*]) \geq 2 \mu(G) \cdot (1 - \eps/4)$ (here $V[M^*]$ stands for the set of endpoints of $M^*$). This implies that there might be at most $\mu(G) \cdot \eps/2$ vertices of $V[M^*]$ which are mapped not mapped to a unique vertex of $V_{\phi^i_j}$ by $\phi^i_j$ amongst other vertices of $V[M^*]$. In turn this implies that $\mu(G) \cdot (1-\eps)$ edges of $M^*$ have both their endpoints mapped to unique vertices of $V_{\phi^i_j}$ by $\phi^i_j$ amongst other endpoints of $\mu(G)$ hence $\mu(G_{\phi_i^j}) \geq \mu(G) \cdot (1-\eps)$.

Observe that this argument holds regardless of the choice of $M^*$ the statement remains true as $G$ undergoes updates even when the updates are made by an adaptive adversary. The contractions $\phi^i_j$ are fixed at initialization. The task of the algorithm is to maintain maximum matching size estimate $\hat{\mu(G)}$ and hence maintain the accurate guess of $\mu(G)$ and to update the contracted graphs. Each contracted graph may undergoes a single update per update to $G$ and there are $\tilde{O}(1)$ contracted graphs. All parts included the worst-case update time of the algorithm is $\tilde{O}(1)$.

%% file: C-trivial-algo.tex
\section{Dynamic $(1+\epsilon)$-Approximate  Matching in $O(n)$ Update Time}
\label{sec:trivial}

Consider the following folklore algorithm for explicitly maintaining a matching: recompute a $(1+\epsilon)$-approximate matching $M$ from scratch in $O(m\epsilon^{-1}\log\epsilon^{-1})$ time \cite{duan2014linear} every after $\epsilon m/2n$ edge updates. Before recomputation, if any edge of $M$ is deleted from the graph, we  delete it from $M$. 

The amortized update time is clearly $\frac{O(m\epsilon^{-1}\log\epsilon^{-1})}{\epsilon m/2n}=O(n\epsilon^{-2}\log\epsilon^{-1})=O(n)$. Also, we have $|M|\ge\mu(G)/(1+\epsilon)-\epsilon m/2n$ where the term $\epsilon m/2n$ is because we might decrease the size of $M$ by $\epsilon m/2n$ before we recompute a new $(1+\epsilon)$-approximate matching. But since $\mu(G) \ge m/ 2n$, we have
\[
|M|\ge\mu(G)/(1+\epsilon)-\epsilon\mu(G)\ge\mu(G)/(1+O(\epsilon))
\]
implying that $M$ is always a $(1+O(\epsilon))$-approximate matching. 

To see why $\mu(G) \ge m/ 2n$, consider the process where we repeatedly choose an edge $e$ and delete both endpoints of $e$ from the graph until no edge is left. Since the set of deleted edges forms a matching, we may repeat at most $\mu(G)$ times. Also, each deletion removes at most $2\Delta$ edges from the graph. Therefore, $m\le\mu(G)\cdot2\Delta\le\mu(G)\cdot2n$.

%% file: table.tex
\section{Tables}

\begin{table}[H]
\footnotesize{
\begin{centering}
\begin{tabular}{|>{\raggedright}p{0.1\textwidth}|>{\centering}p{0.15\textwidth}|>{\centering}p{0.1\textwidth}|>{\centering}p{0.1\textwidth}|>{\centering}p{0.1\textwidth}|>{\centering}p{0.15\textwidth}|>{\centering}p{0.1\textwidth}|}
\hline 
Model & \multicolumn{2}{c|}{Adjacency List } & \multicolumn{2}{c|}{Adjacency List } & \multicolumn{2}{c|}{Adjacency Matrix}\tabularnewline
\hline 
Guarantee & Approx & Time & Approx & Time & Approx & Time\tabularnewline
\hline 
\hline 
\cite{parnas2007approximating} & $(2,\epsilon n)$ & $\Delta^{O(\log(\Delta/\epsilon))}$ &  &  &  & \tabularnewline
\hline 
\multirow{2}{0.1\textwidth}{\cite{nguyen2008constant}} & $(2,\epsilon n)$ & $2^{O(\Delta)}/\epsilon^{2}$ &  &  &  & \tabularnewline
\cline{2-7} 
 & $(1,\epsilon n)$ & $2^{\Delta^{O(1/\epsilon)}}$ &  &  &  & \tabularnewline
\hline 
\multirow{2}{0.1\textwidth}{\cite{yoshida2009improved}} & $(2,\epsilon n)$ & $\Delta^{4}/\epsilon^{2}$ &  &  &  & \tabularnewline
\cline{2-7} 
 & $(1,\epsilon n)$ & $\Delta^{O(1/\epsilon^{2})}$ &  &  &  & \tabularnewline
\hline 
\cite{onak2012near,chen2020sublinear} & $(2,\epsilon n)$ & $(d+1)\Delta/\epsilon^{2}$  &  &  & $(2,\epsilon n)$ & $n\sqrt{n}/\epsilon^{2}$ \tabularnewline
\hline 
\cite{behnezhad2022time} & $(2,\epsilon n)$ & $(d+1)/\epsilon^{2}$ & $2+\epsilon$ & $n+\Delta/\epsilon^{2}$ & $(2,\epsilon n)$ & $n/\epsilon^{3}$\tabularnewline
\hline 
\cite{behnezhad2023beating} & $(2-\frac{1}{2^{O(1/\gamma)}},o(n))$ & $(d+1)\Delta^{\gamma}$ & $2-\frac{1}{2^{O(1/\gamma)}}$ & $n+\Delta^{1+\gamma}$ & $(2-\frac{1}{2^{O(1/\gamma)}},o(n))$ & $n^{1+\gamma}$\tabularnewline
\hline 
\cite{bhattacharya2022sublinear,behnezhad2022sublinear} & $(1.5,\epsilon n)$ & $nd^{1-\Omega(\epsilon^{2})}$ & $1.5+\epsilon$ & $n\Delta^{1-\Omega(\epsilon^{2})}$ & $(1.5,\epsilon n)$ & $n^{2-\Omega(\epsilon^{2})}$\tabularnewline
\hline 
\cite{behnezhad2022sublinear}\\
bipartite graph only & $(1.5-\Omega(1),o(n))$ & $n^{2-\Omega(1)}$ & $1.5-\Omega(1)$ & $n^{2-\Omega(1)}$ & $(1.5-\Omega(1),o(n))$ & $n^{2-\Omega(1)}$\tabularnewline
\hline 
\textbf{Our} & &  &  &  & $(1,\eps n)$ & $n^{2-\Omega_{\eps}(1)}$\tabularnewline
\hline 
\end{tabular}
\par\end{centering}
}

\caption{\label{tab:size}Summary of sublinear-time algorithms for estimating the size of maximum matching. We omit $\protect\polylog(n/\epsilon)$ factors. $\Delta$ and $d$ denote the maximum and average degree of the graph, respectively.}

\end{table}

\begin{table}[H]
\begin{centering}
\footnotesize{%
\begin{tabular}{|>{\raggedright}m{0.2\textwidth}|>{\centering}p{0.2\textwidth}|>{\centering}p{0.15\textwidth}|}
\hline 
Reference & Approximation & Query time\tabularnewline
\hline 
\hline 
\cite{parnas2007approximating,marko2009approximating} & $2+\epsilon$ & $\Delta^{O(\log(\Delta/\epsilon))}$\tabularnewline
\hline 
\cite{rubinfeld2011fast,alon2012space} & $2$ & $\Delta^{O(\Delta\log\Delta)}$\tabularnewline
\hline 
\cite{reingold2016new} & $2$ & $2^{O(\Delta)}$\tabularnewline
\hline 
\multirow{3}{0.2\textwidth}{\cite{levi2015local}} & $2$ & $\Delta^{O(\log^{2}\Delta)}$\tabularnewline
\cline{2-3} 
 & $2+\eps$ & $\Delta^{4}$\tabularnewline
\cline{2-3} 
 & $1+\eps$ & $\Delta^{O(1/\eps^{2})}$\tabularnewline
\hline 
\cite{ghaffari2016improved} & $2$ & $\Delta^{O(\log\Delta)}$\tabularnewline
\hline 
\cite{ghaffari2019sparsifying} & $2$ & $\Delta^{O(\log\log\Delta)}$\tabularnewline
\hline 
\cite{ghaffari2022local} & $2$ & $\Delta^{O(1)}$\tabularnewline
\hline 
\cite{kapralov2020space} & $O(1)$ in expectation & $\Delta$\tabularnewline
\hline 
\textbf{Our} & $(1,\eps n)$ & $n^{2-\Omega_{\eps}(1)}$\tabularnewline
\hline 
\end{tabular}}
\par\end{centering}
\caption{\label{tab:LCA}Summary of local computation algorithms for matching oracles. We omit $\protect\polylog(n/\epsilon)$ factors. All 2-approximation algorithms actually compute a maximal independent set. $\Delta$ and $d$ denote the maximum and average degree of the graph, respectively.}
\end{table}